\documentclass[useAMS,usenatbib]{mn2e}

\usepackage{graphicx}
\usepackage{amssymb}
\usepackage{multirow}
\usepackage{pifont}
\usepackage{longtable}
\usepackage{bbold}
\usepackage{color}

\makeatletter
    
    \newcommand{\Rmnum}[1]{\expandafter\@slowromancap\romannumeral #1@}
\makeatother

\title[Inclined KBOs in the high-order resonances]
{A study of the high-inclination population in the Kuiper belt -- \Rmnum4. High-order mean motion resonances in the classical region}
\author[Jian Li, S. M. Lawler and Hanlun Lei]
{Jian Li$^{1,2}$\thanks{E-mail: ljian@nju.edu.cn}, S. M. Lawler$^3$ and Hanlun Lei$^{1,2}$\\
$^1$School of Astronomy and Space Science, Nanjing University, 163 Xianlin Avenue, Nanjing 210023, PR China\\
$^2$ Key Laboratory of Modern Astronomy and Astrophysics in Ministry of Education, Nanjing University, Nanjing 210023, PR China\\
$^3$Campion College, University of Regina, Regina, SK S4S 0A2, Canada}
\begin{document}

\date{Accepted 1988 December 15. Received 1988 December 14; in original form 1988 October 11}

\pagerange{\pageref{firstpage}--\pageref{lastpage}} \pubyear{2002}

\maketitle

\label{firstpage}

\begin{abstract}

In our previous study of Neptune's 4:7 mean motion resonance (MMR), we discovered that its resonant angle can only librate within a specific eccentricity ($e$) versus inclination ($i$) region, determined by a theoretical limiting curve \citep{Li2020}. This ``permissible region'' is independent of time and encompasses the entire possible stable region. We now generalize this theory to investigate all high-order MMRs embedded in the main classical Kuiper belt (MCKB). We first consider the 2nd-order 3:5 MMR in the framework of planet migration and resonance capture, and have further validated our limiting curve theory for both captured and observed 3:5 resonators. It suggests that only the $(e, i)$ pairs inside the individual permissible regions should be chosen as initial conditions for studying the in-situ evolution of high-order resonators. With such a new setting, we proceed to explore the long-term stability (for 4 Gyr) of different resonant populations, and our simulations predict that: (1) the 3:5 and 4:7 resonators are comparable in number, and they could have inclinations up to $40^{\circ}$; (2) the populations of objects in the higher order 5:9, 6:11, 7:12 and 7:13 resonances is about 1/10 of the 3:5 (or 4:7) resonator population, and nearly all of them are found on the less inclined orbits with $i<10^{\circ}$; (3) for these high-order resonances, almost all resonators reside in their individual permissible regions. In summary, our results make predictions for the number and orbital distributions of potential resonant objects that will be discovered in the future throughout the MCKB.

 \end{abstract}


\section{Introduction}

\begin{figure}
 \centering
  \begin{minipage}[c]{0.5\textwidth}
  \hspace{-1 cm}
  \centering
  \includegraphics[width=8.6cm]{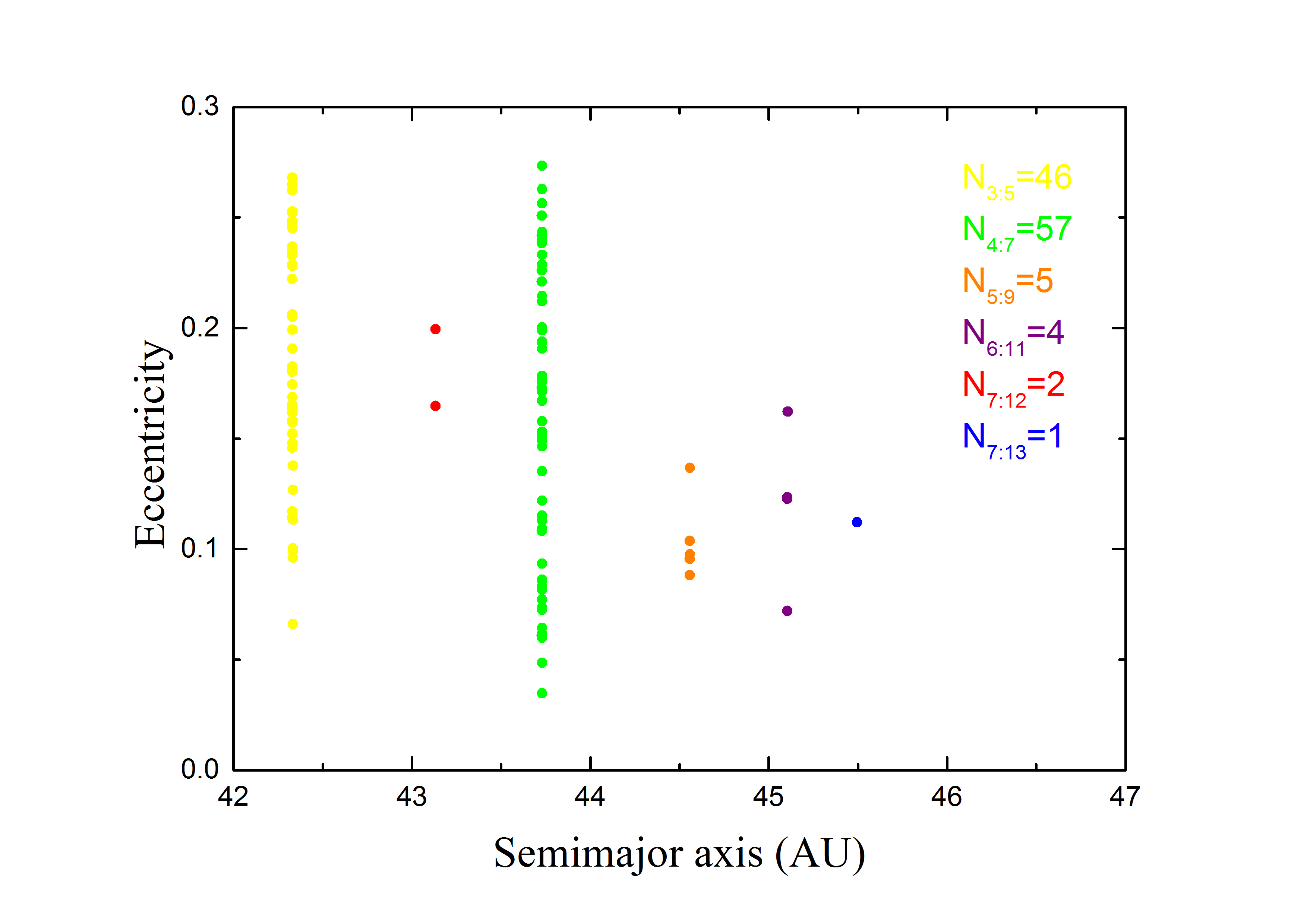}
  \end{minipage}
  \begin{minipage}[c]{0.5\textwidth}
  \hspace{-1 cm}
  \centering
  \includegraphics[width=8.6cm]{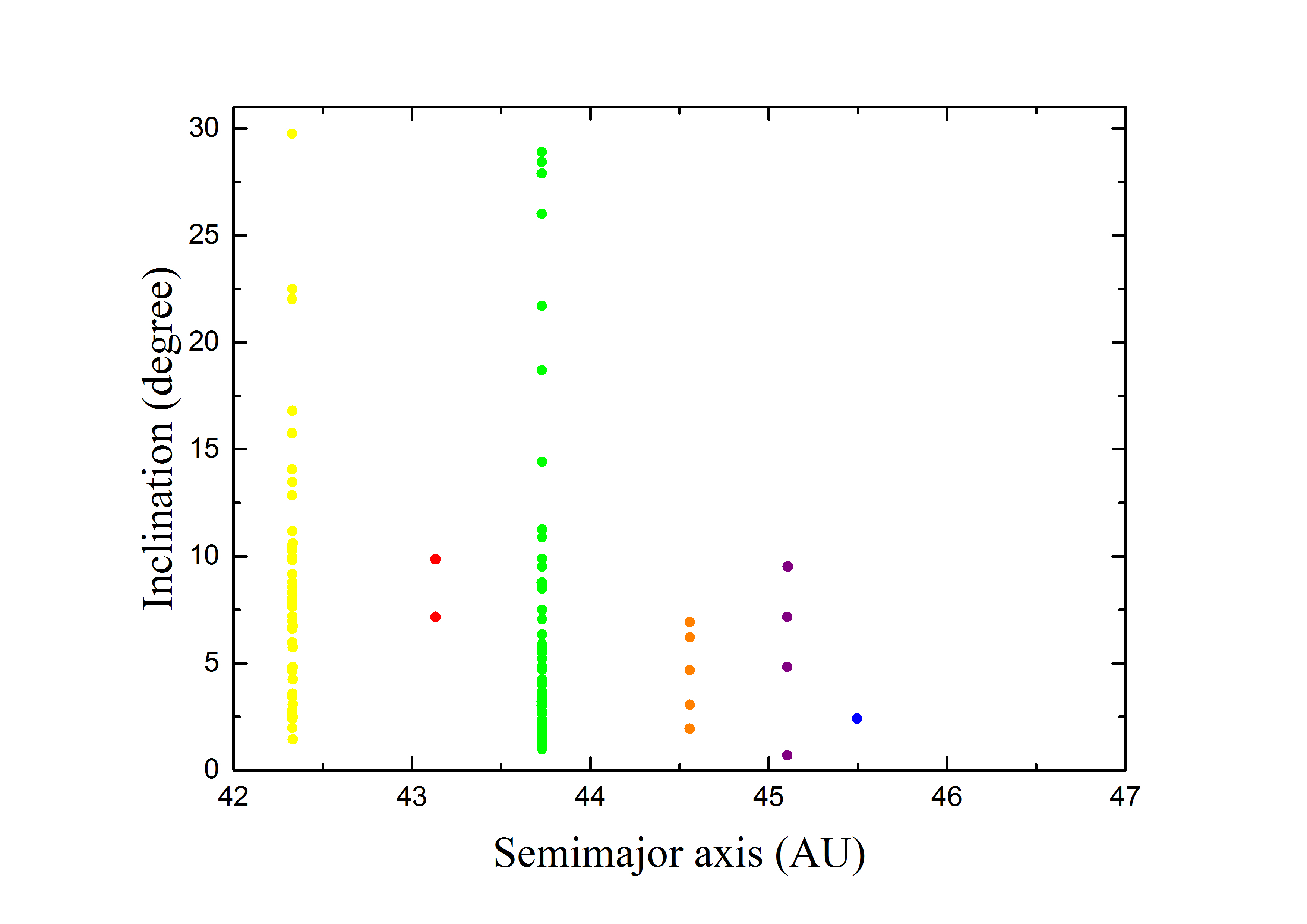}
  \end{minipage}
  \caption{The orbital distributions of observed KBOs identified in the 2nd-order and higher-order resonances within the MCKB region between 42 and 47 AU. From smaller to larger semimajor axis, these KBOs occupy the 3:5 (yellow), 4:7 (green), 5:9 (orange), 6:11 (purple), 7:12 (red), and 7:13 (blue) resonances.  The data are taken from the Minor Planet Center as of May 13, 2020. Only KBOs with multiple-opposition observation arcs are shown, with very small uncertainties characterised by their measured semimajor axes $\Delta a/a<0.16\%$ (see Table \ref{realRKBOlist}).}
  \label{real}
\end{figure}

The Kuiper belt consists of numerous icy celestial bodies beyond the orbit of Neptune in the outer Solar system. To provide new insights into the formation and evolution of those bodies with high orbital inclinations, we have carried out a series of works entitled ``A study of the high-inclination population in the Kuiper belt''. In the first two papers, our study began by exploring the dynamics of objects in the 2:3 \citep[\Rmnum1. Plutinos,][]{Li2014a} and 1:2 \citep[\Rmnum2. Twotinos,][]{Li2014b} mean-motion resonances in the Kuiper belt. After thoroughly analyzing these two first-order resonances, the third paper proceeded to examine a typical high-order resonance \citep[\Rmnum3. The 4:7 mean-motion resonance,][]{Li2020}. In this fourth work (i.e. paper \Rmnum4), we expand our investigations to include all the high-order resonances in the main classical Kuiper belt (MCKB).

The MCKB extends between roughly 42 AU and 47 AU, where a large fraction of objects have small-eccentricity and low-inclination orbits \citep{Glad2008, Peti2011}. Embedded in the MCKB region, there are a number of typically higher eccentricity objects librating within the high-order mean-motion resonances (MMRs) with Neptune. In order of increasing resonance order, the occupied resonances include the 3:5, 4:7, 5:9, 6:11, 7:12, and the 7:13. Generally, as the order of a resonance becomes higher, the resonance's strength is weaker, and the overlap with its adjacent resonances may induce instability. A comprehensive study of the Kuiper Belt Objects (KBOs) in these various MMRs could provide information about the process of planet migration in the early Solar system \citep[e.g.][]{Pike2017, Pira2021}.

The dynamics of the relatively low-order 3:5 and 4:7 MMRs have been well investigated  \citep{Meli2000, Glad2012, Gall2018, Gall2020}, and over 100 resonant KBOs (RKBOs) associated with these two resonances have been observed so far. As more KBOs were discovered in the early 2000s, it became clear that the community needs better understanding of the 5:9, 6:11, 7:12 and even higher order resonances \citep{Chia2003, Lyka2005}. \citet{Lyka2007} and \citet{Glad2008} classified the KBOs observed at that time, they identified two objects (2002 GD32 and 2001 KU76) locked in the 5:9 MMR and one (2001 KU76) in the 6:11 MMR. \citet{Lyka2007} also conducted numerical simulations to check the long-term evolution of the 5:9 resonators, showing that the stability limits are of eccentricities $e<0.185$ and inclinations $i<15^{\circ}$, which are consistent with the observed orbits of 2002 GD32 and 2001 KU76. The Outer Solar System Origins Survey (OSSOS) provided dynamical classifications for all characterized discoveries \citep{Bann2018} and found many RKBOs, including 14 in the 3:5, one in the 5:9, two in the 11:6, and one in the 7:12. Very recently, \citet{Khai20} carried out dynamical classification of 240 KBOs detected by the Dark Energy Survey, and they identified 2013 TB172 to be a 7:12 resonator.

Fig. \ref{real} shows the eccentricity (upper panel) and inclination (lower panel) distributions of the currently observed KBOs classified as resonators, from the 2nd-order 3:5 MMR to the 6th-order 7:13 MMR. The identification of RKBOs follows the similar procedure used in our previous works \citep{Li2014a, Li2014b, Li2020}. First, we select the KBOs with semimajor axes $a=42$-47 AU registered in the IAU Minor Planet Center\footnote{https://minorplanetcenter.net/iau/lists/MPLists.html}, as of May 13, 2020, and only the samples with observations at two or more oppositions are considered. Then the KBOs' orbits are numerically integrated for a time-span of $10^8$ yr, including  the gravitational perturbations of four Jovian planets. At the end of the integration, we examine the time evolution of the resonant angles of various MMRs, and an object is regarded as a resonator if some resonant angle librates throughout the last 10 Myr of the integration. We can see in Fig. \ref{real} that there are tens of KBOs located inside either the 3:5 MMR or the 4:7 MMR; while the numbers of resonant populations decrease sharply for the 5:9, 6:11, 7:12, and 7:13 MMRs. More information about the identified RKBOs is provided in Appendix A. 

We additionally find in the MPC database that, no RKBOs are observed in the 7-th order 8:15, 9:16, 10:17, 11:18 MMRs or even higher order resonances. This may be due to the fact that, as the resonance order increases, the resonance strength further decreases and thus we expect fewer KBOs in these resonances. \citet{volk2016} reported dozens of RKBOs detected in the OSSOS Survey. They argued that the 7th-order resonances may hold some KBOs, such as 2013 HR156 in the 8:15 MMR and 2013 GU136 in the 9:16 MMR, but their orbits are insecure (meaning that the three tested orbits within the measured orbital uncertainties did not all librate in the resonance for a full 10~Myr). We do not find any RKBOs librating in 7th or higher order resonances, likely because we adopt the best-fit orbits listed in the MPC database.

Focusing on the formation of RKBOs, a few previous works have been conducted to investigate the effects of Neptune's migration. In the framework of a Nice model migration simulation including a dynamical instability of the giant planets from \citet{bras2013}, \citet{Pike2017} and \citet{PL2017} explored the resulting distribution of RKBOs. At the end state of their simulations, many stable 3:5 and 4:7 resonators are present, but there are only a few stable objects trapped in other high-order resonances within the MCKB region. There could be two possible reasons for this outcome: (1) the capture/survival rates of objects in the higher order resonances (e.g. 5:9, 6:11, 7:13 MMRs) are quite low, one may need to increase the number density of test particles in such simulations; (2) the considered migration process could be too violent to maintain objects in the resonances of 4th- or higher orders. \citet{lawl19} performed a similar analysis for simulations from \citet{kaib16}, which included grainy and smooth Neptune migrations with different timescales. This analysis found many stable 3:5 and 4:7 resonators with different frequencies depending on Neptune's migration speed and mode; but due to the limitation of a relatively small number of test particles, higher order resonances were not examined.

Beyond their dynamical behaviors, RKBOs can also be characterized by the physical properties such as surface colors. \citet{shep2012} analyzed the optical surface colours of 58 KBOs in Neptune's MMRs, and they found that various resonant populations could have significantly different colour distributions. It is interesting to note that, for the 3:5 and 4:7 resonant objects, their colours are mostly ultra-red, similar to the low-inclination non-resonant objects in the MCKB, but different from the resonant objects interior (e.g. the 3:4 and 2:3 MMRs) or exterior (e.g. the 1:2 and 2:5 MMRs) to the MCKB. In addition, \citet{pike2022} measured surface colours for 92 KBOs from the OSSOS dataset, including several KBOs in high-order (e.g. 3:5, 4:7, 5:9, and 6:11) resonances in the MCKB region. They confirmed the presence of the same colour (faint infrared class) surfaces in all these resonances. These observational results imply that the MCKB resonant populations may share a common origin and evolutionary path. Therefore, it is important that we investigate these populations all together.

In this paper, we build on the work of \citet{Li2020} (hereafter MN2020), which studied the high-inclination population in the 4:7 MMR in detail. They completed a detailed analysis of the two different resonant modes, the eccentricity-type and the mixed eccentricity-inclination-type, while the latter appears only in highly inclined orbits. Both the observed and simulated 4:7 populations are dominated by the eccentricity-type resonance. To better describe the eccentricity-type resonators with high inclinations, \citet{Li2014a} defined two kinds of libration centers of the resonant angle: the special libration center (SLC) and the general libration center (GLC). The SLC corresponds to the stable equilibrium point where the time variation of the semimajor axis satisfies ${\rm d}a/{\rm d}t=0$. Depending on the orbital eccentricity $e$ and inclination $i$, they demonstrate that the SLC oscillates when the argument of perihelion varies between 0 and $360^{\circ}$, and the oscillation amplitude of the SLC determines the lower limit of the resonant amplitude, denoted by $A_{\min}$. The mean value of the SLC during one period of $\omega$ is defined as the GLC, which actually corresponds to the libration center of the resonant angle. The symmetric resonances (i.e. not the 1:$n$ type) have just one stable libration centre at $180^{\circ}$, and the libration of the resonant angle is allowed under the condition of $A_{\min}<180^{\circ}$. For the first-order 2:3 resonance, $A_{\min}$ has a maximum value of $\sim75^{\circ}$, so libration is always possible for any $(e, i)$ pair. However, MN2020's investigation of the high-order 4:7 resonance revealed an intriguing finding for high inclination resonators: For a given value of $i$, $A_{\min}$ can only be limited to $<180^{\circ}$ for orbits with $e$ exceeding a critical value $e_c$. They showed $e_c$ increases with inclination $i$, and thus introduced a `limiting curve' $e_c=e_c(i)$ on the $(e, i)$ plane. In the region on the left side of this curve (i.e. $e\ge e_c$), the condition $A_{\min}<180^{\circ}$ is fulfilled and thus the resonant angle is permitted to librate. Accordingly, this region is referred to as the `permissible region' of the 4:7 resonators. We suppose that the concept of the limiting curve is an important consideration both for the analysis of observed RKBOs and for initializing simulations that explore dynamical evolution of resonators in the 4:7 and other high-order MMRs within the MCKB. Below we will focus on the eccentricity-type high-order MMRs in the region of the MCKB. These resonances all have their libration centers of the critical resonant angles located at $180^{\circ}$, unlike the 1:$n$ resonances which additionally have asymmetric libration centers \citep{Li2014b}.

To gain a deep understanding of the high-order MMRs embedded in the MCKB, our aim is to achieve the following goals:\\ 
$~~~(\mathbb 1)$ We want to reveal the global dynamics of the high-order resonances, especially for the orbits with high inclinations. To acheive this, we further develop our limiting curve theory in order to study the dynamical behaviours of resonant particles in other high-order resonances.\\ 
$~~~(\mathbb 2)$ The number of currently observed RKBOs in the 4:7 resonance is only 57, but the intrinsic population with $Hr<$ 8.66 (approximately corresponding to diameters of $>100$~km) could be as many as 1000 \citep{volk2016}. Therefore, we want to estimate whether a substantial number of undetected objects could reside in other high-order resonances, by evaluating the survival rates of different resonant populations in the long-term evolution simulations.\\
$~~~(\mathbb 3)$ We would like to use long-term simulations produce population and distribution estimates for the escaped resonators. These objects may evolve to classical (non-resonant) KBOs with small eccentricities and high inclinations. The deficit of classical KBOs on such orbits produced in Nice model simulations is an issue yet to be resolved \citep{Levi2008}.

The rest of this paper is organised as follows: in Section 2, we revisit the simulations from MN2020 and apply our limiting curve theory to predict the orbital distribution of the 3:5 RKBOs. In Section 3, we investigate the phase space structures of individual resonances ranging from the 2nd- to 7th-order, using the planar circular restricted 3-body problem model to calculate the widths of these resonances on the $(a, e)$ plane. In Section 4, given specific initial conditions constrained by the limiting curve theory, we explore the long-term evolution of objects originally trapped in six different high-order resonances. The long-term survivors have powerful implications for the phase space where we expect to discover RKBOs in the future, as well as provide predictions for non-resonant classical KBOs that should be present today on nearly-circular, highly-inclined orbits. Finally, the conclusions and discussion are given in Section 5.


\section[]{3:5 resonators from migration simulations}


\begin{figure}
 \hspace{0cm}
  \centering
  \includegraphics[width=9cm]{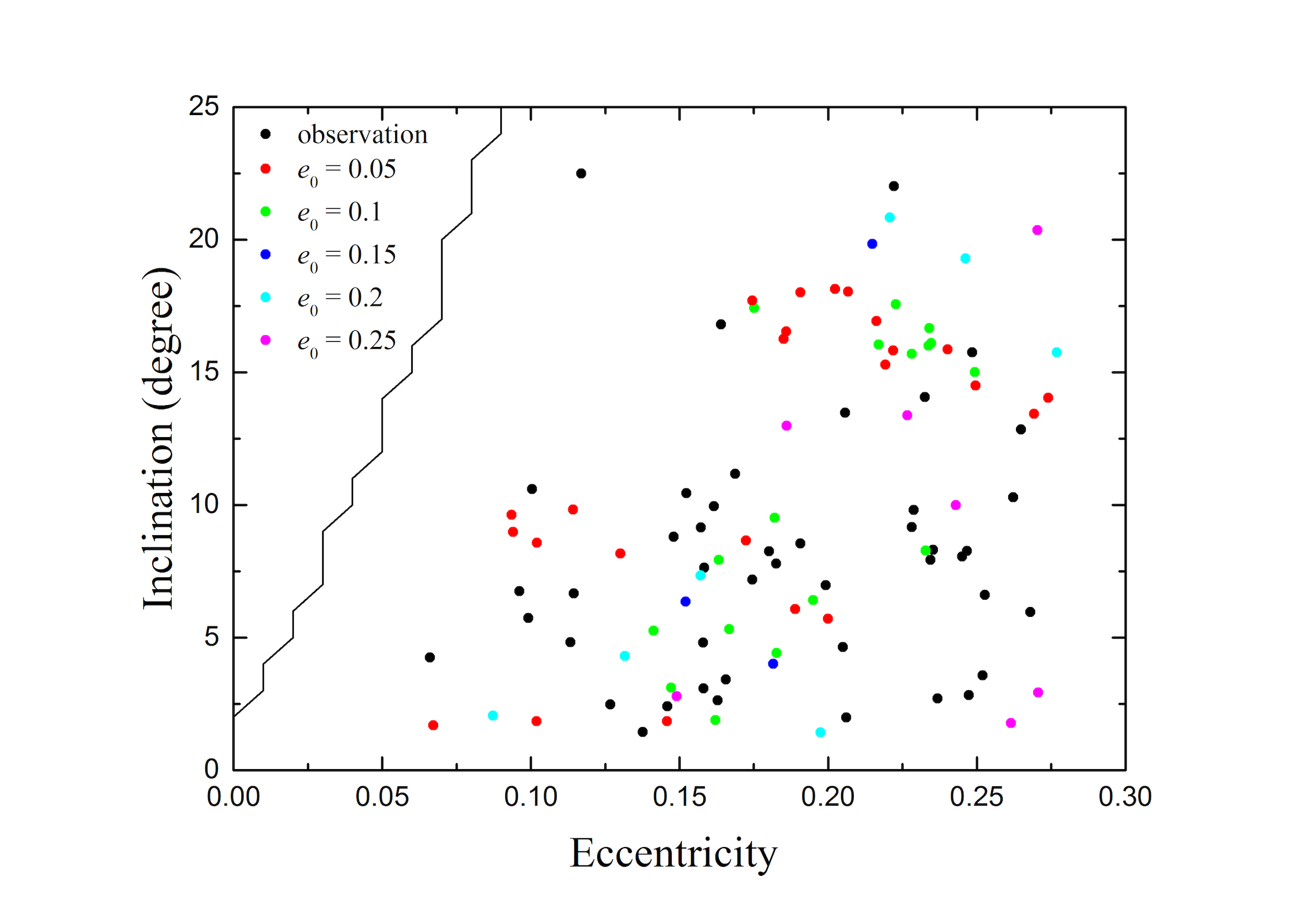}
  \caption{A scatter plot of eccentricities and inclinations for the observed 3:5 RKBOs (black dots), and the simulated particles captured into the 3:5 MMR (colourful dots) after the sweeping of the 4:7 MMR in the planet migration model constructed in MN2020. All these 3:5 resonators are associated to the eccentricity-type resonance, and without exception, they reside in the permissible region to the right of the limiting curve (black line). For the simulated resonators, each colour represents a subset with a specific initial eccentricity of $e_0\in[0.05, 0.25]$ (see legend within figure). Each subset includes a range of initial inclinations from $1^{\circ}$ to $30^{\circ}$.}
  \label{simulated35}
\end{figure}

As we showed in MN2020, the distribution of eccentricities and inclinations of captured 4:7 resonators at the end of planet migration simulations are confined by the limiting curve. This suggests that the initial ($e, i$) pairs for the usual long-term evolution simulations do not need to be uniformly distributed across all possible values, but should be restricted to the permissible region, i.e. on the right side of the associated limiting curve for the eccentricity-type resonance.


To provide a more compelling demonstration, here we conduct an additional investigation of the 3:5 resonators generated in the MN2020 Neptune migration simulations. Before delving into the details, we first briefly review the principal settings of our migration simulations: (1) as in the classical migration scenario proposed by \citet{Malh1995}, Neptune starts from 28.2 AU and moves smoothly outwards to its current position at 30.2 AU. The migration time-scale is chosen to be $2\times10^7$ yr to maintain the adiabatic invariant condition \citep{Meli2000}. (2) Test particles are distributed in the region $41.9\le a \le42.4$ AU, which is within the area of the MCKB and can be swept through by both the 4:7 and 3:5 MMRs. The initial eccentricities of the particles range from 0.05 to 0.25, with increments of 0.05; and the initial inclinations are set to be $1^{\circ}$, $10^{\circ}$, $20^{\circ}$, and $30^{\circ}$. (3) The system is integrated for $10^8$ yr, which is sufficiently long for Neptune to reach its current orbit and evolve there without migration for several tens of Myr.

In the framework of the Neptune migration model described above, we find that after the 4:7 MMR has swept over the test particle disk, some residual particles are picked up as the 3:5 MMR sweeps through afterwards. Although in this simulation the 3:5 MMR only passes through a very narrow disk extending from 41.9 to 42.4 AU, the results may have some substantial implications for the observed 3:5 RKBOs. As a 2nd-order resonance, the 3:5 MMR has two resonant angles:
\begin{eqnarray}
&&\!\!\!\!\!\!\!\!\!\!\!\!\sigma_e=5\lambda-3\lambda_N-2\varpi,\nonumber\\
&&\!\!\!\!\!\!\!\!\!\!\!\!\sigma_i=5\lambda-3\lambda_N-2\Omega,
\label{angle35}
\end{eqnarray}
where $\sigma_e$ and $\sigma_i$ characterize the eccentricity- and inclination-type resonances, respectively. For the KBO, $\lambda$ is the mean longitude, $\varpi$ is the longitude of perihelion, $\Omega$ is the longitude of ascending node; and $\lambda_N$ refers to the mean longitude of Neptune. The stability of the eccentricity-type 3:5 MMR was studied by \citet{Meli2000} for the case of low inclinations ($i=1^{\circ}$). Later \citet{Lyka2007} studied higher inclinations, and they found that most of the simulated stable 3:5 resonators have orbits with $e<0.3$ and $i<20^{\circ}$.  

\begin{figure}
 \hspace{0cm}
  \centering
  \includegraphics[width=9cm]{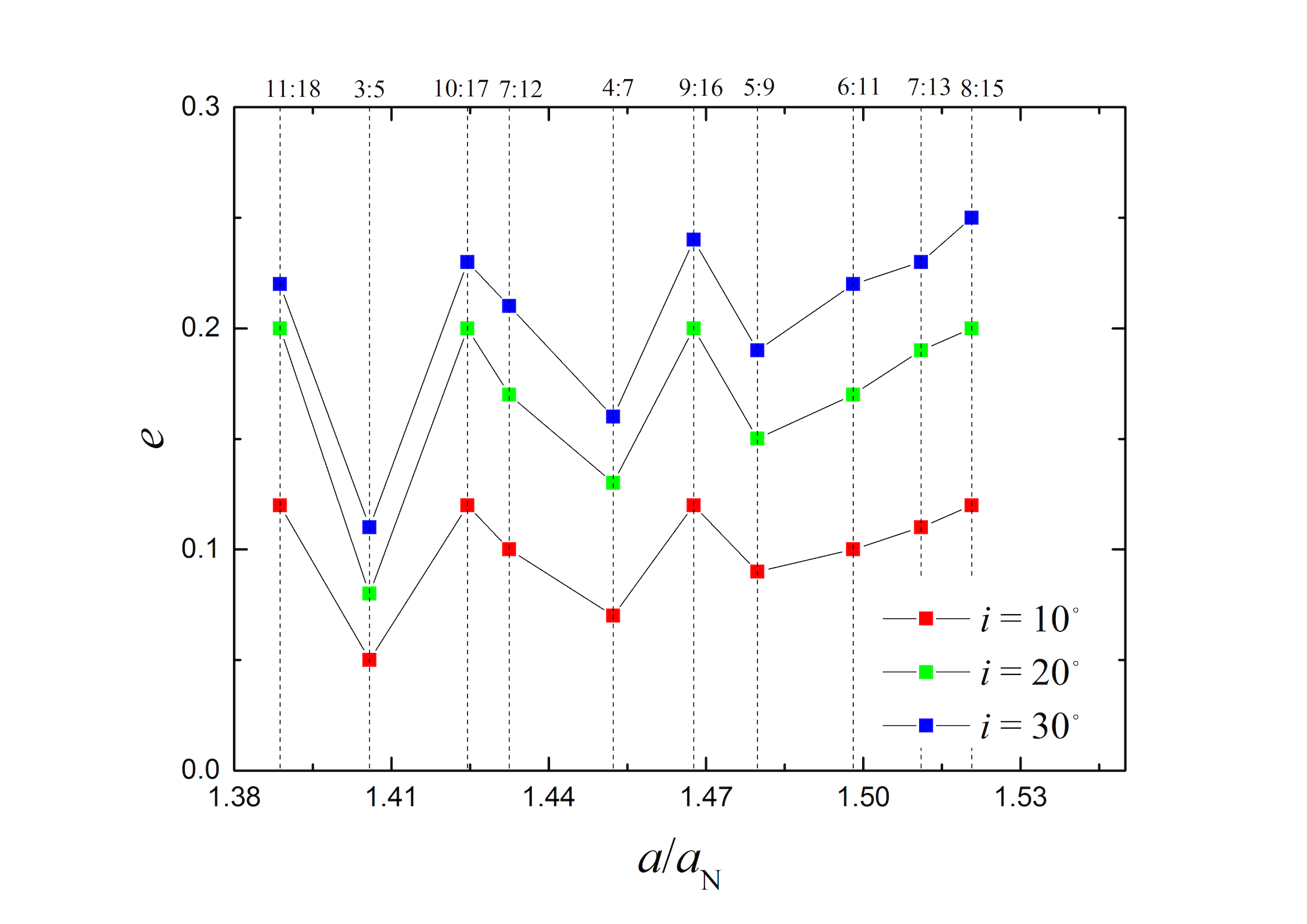}
  \caption{Our prediction for the most likely eccentricities for high-inclination RKBOs in the MCKB, based on where their resonant angles are allowed to librate. Semimajor axis values, $a$, are in respect with Neptune's semimajor axis $a_N$. Dashed lines denote the positions of individual resonances ranging from the 2nd- to 7th-order. The red, green and blue squares correspond to the cases of $i=10^{\circ}$, $20^{\circ}$, and $30^{\circ}$, respectively. The RKBOs at each of these inclinations are expected to have eccentricities larger than the values indicated by the squares.}
  \label{resonances}
\end{figure}

Similar to MN2020's analysis of the 4:7 MMR, Fig. \ref{simulated35} shows the eccentricities and inclinations of the simulated particles captured into the 3:5 MMR (colourful dots). For comparison, the currently observed 3:5 RKBOs from Fig. \ref{real} also have been plotted in this figure (black dots). First, we find that both the simulated and observed populations are associated with the eccentricity-type resonance, and most of them tend to inhabit the stable $e$ and $i$ ranges mentioned above. As we expected, all of these 3:5 resonators indeed lie in the permissible region, on the right side of the limiting curve (black line), where the resonant angle $\sigma_e$ is allowed to librate. Secondly, we find that during Neptune's outward migration, the probability of capturing planetesimals into the inclination-type 3:5 resonance is extremely low. A careful examination of our numerical results shows that although a few particles show libration of the $\sigma_i$ resonant angle, they are actually experiencing the eccentricity-type MMR coupled with the Kozai mechanism. We do not find any independent inclination-type 3:5 resonators in our simulations. 

Combined with previous results for the 4:7 MMR (MN2020), our simulations strongly suggest that the eccentricity-type resonance should be the dominant mode for the 4:7, 3:5, and other higher order MMRs, even for the high-inclination population. Consequently, only the eccentricity-type resonance will be considered for the rest of this paper. 

It is important to note that, for the 4:7 and 3:5 eccentricity-type resonators, higher inclinations require higher eccentricities, due to the constraints of the individual limiting curves, e.g. the $(e, i)$ distribution shown in Fig. \ref{simulated35}. Applying the semi-analytical method developed in MN2020 to the 2nd- to 7th-order MMRs embedded in the MCKB, we can determine a set of lower limits for eccentricities at several representative high inclinations of $i=10^{\circ}, 20^{\circ}$, and $30^{\circ}$, as depicted in Fig. \ref{resonances}. The undetected high-order RKBOs on inclined orbits are expected to be in the $e$ range above the individual squares. Notice that for these highly inclined RKBOs, the higher the resonance order is, the larger their eccentricities become, and the resulting smaller perihelia should make discovery easier. 



\section{Widths of high-order resonances in the MCKB} \label{sec:dyn}

\begin{figure*}
  \centering
  \begin{minipage}[c]{1\textwidth}
  \vspace{0 cm}
  \includegraphics[width=9cm]{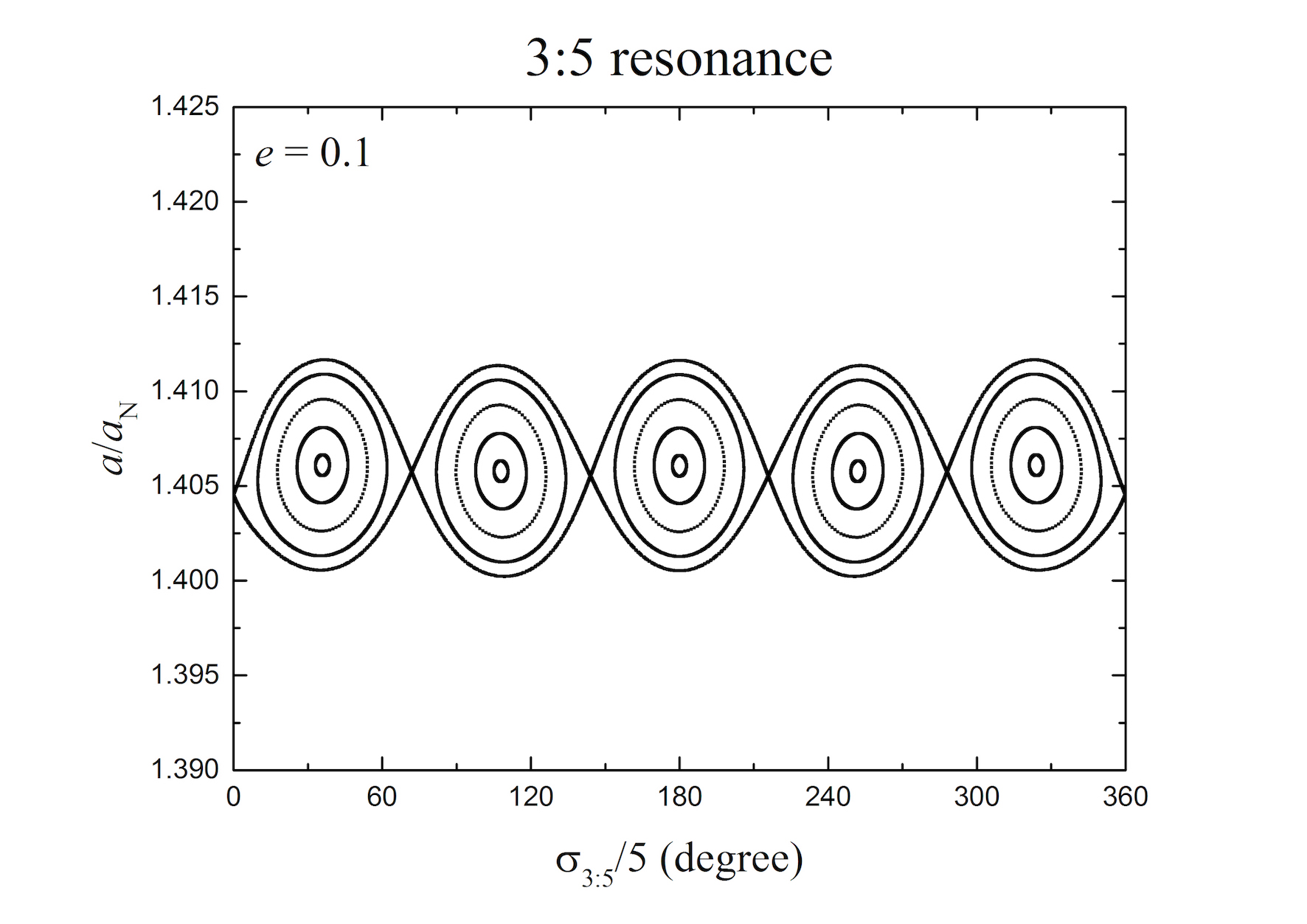}
  \includegraphics[width=9cm]{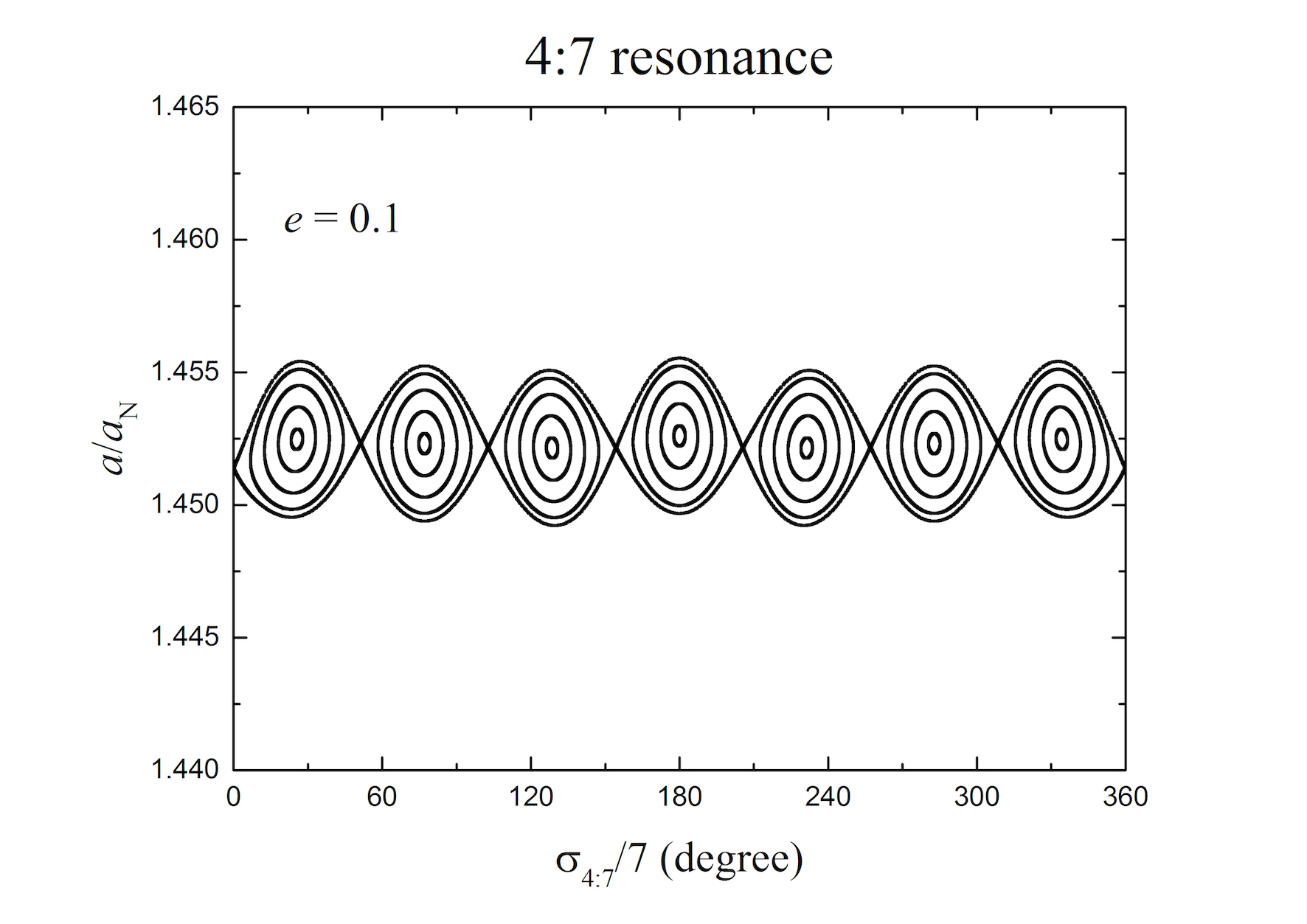}
  \end{minipage}
  \begin{minipage}[c]{1\textwidth}
  \vspace{0 cm}
  \includegraphics[width=9cm]{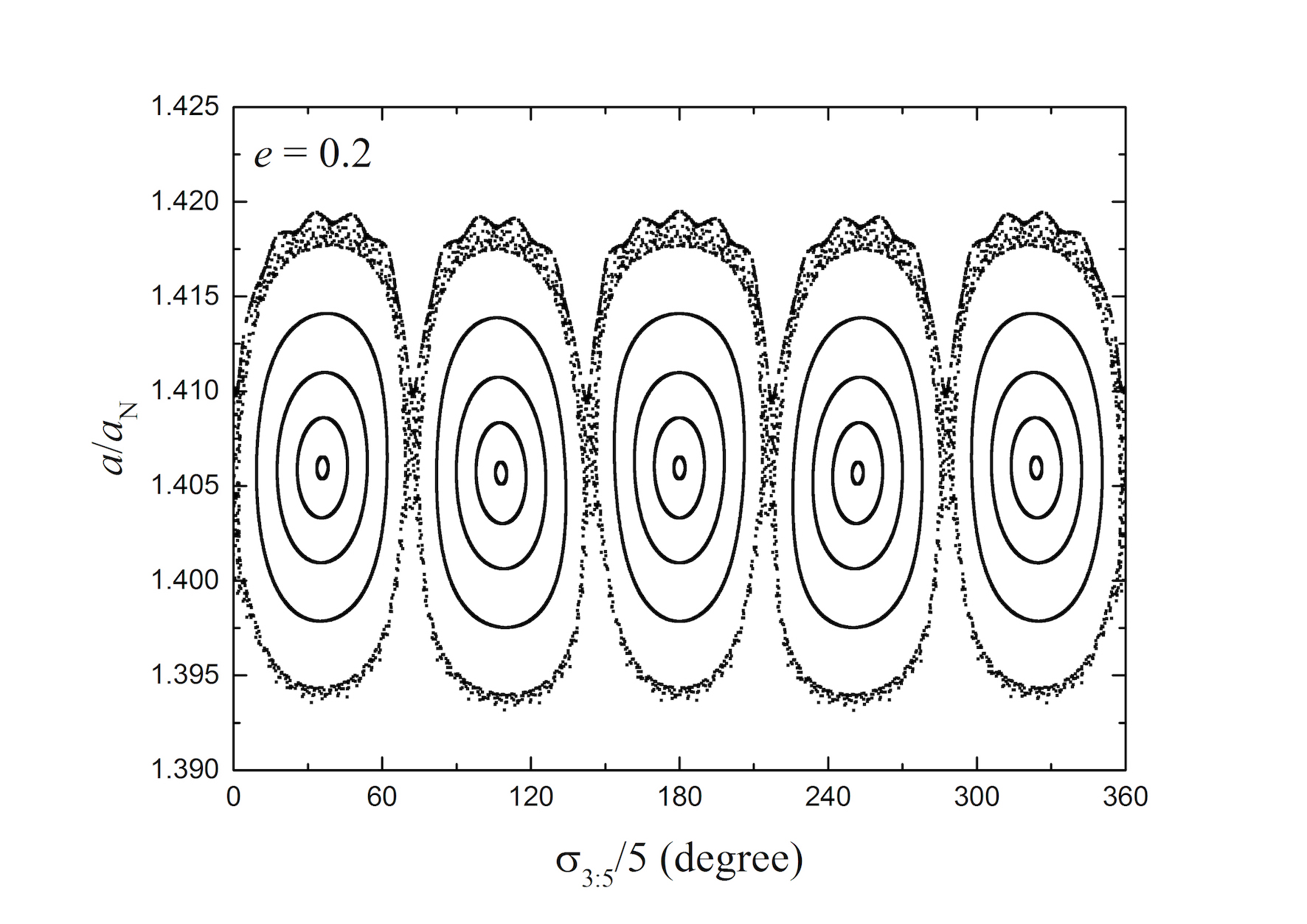}
  \includegraphics[width=9cm]{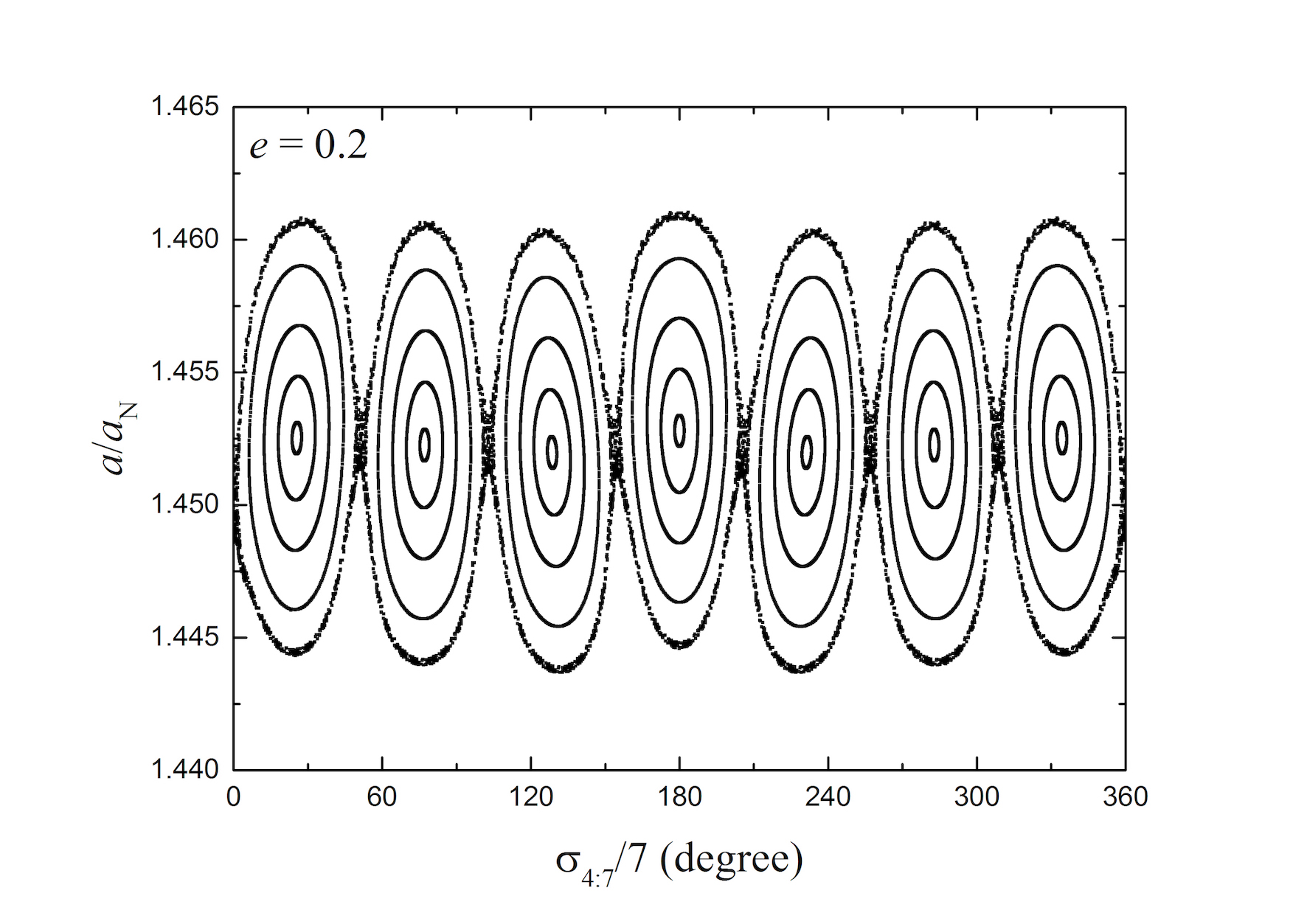}
  \end{minipage}
  \begin{minipage}[c]{1\textwidth}
  \vspace{0 cm}
  \end{minipage}
  \caption{Examples of the phase spaces of the high-order resonances in the MCKB region, determined by the planar circular restricted 3-body problem model. Left panels are for the 3:5 resonance at $e=0.1$ (top) and $e=0.2$ (bottom), and right panels show the same for the 4:7 resonance. For each $p$:$q$ resonance, we adopt a new critical resonant angle of $\sigma_{p:q}/q$, where $\sigma_{p:q}$ is the usual resonant critical angel defined by equation (\ref{ResAng}). In this way, $q$ resonant islands can be seen clearly. The vertical axis is the resonant particles' semimajor axis $a$, which is scaled by Neptune's semimajor axis $a_N$.} 
 \label{Pspace}
\end{figure*} 

\begin{figure}
 \hspace{0cm}
  \centering
  \includegraphics[width=9cm]{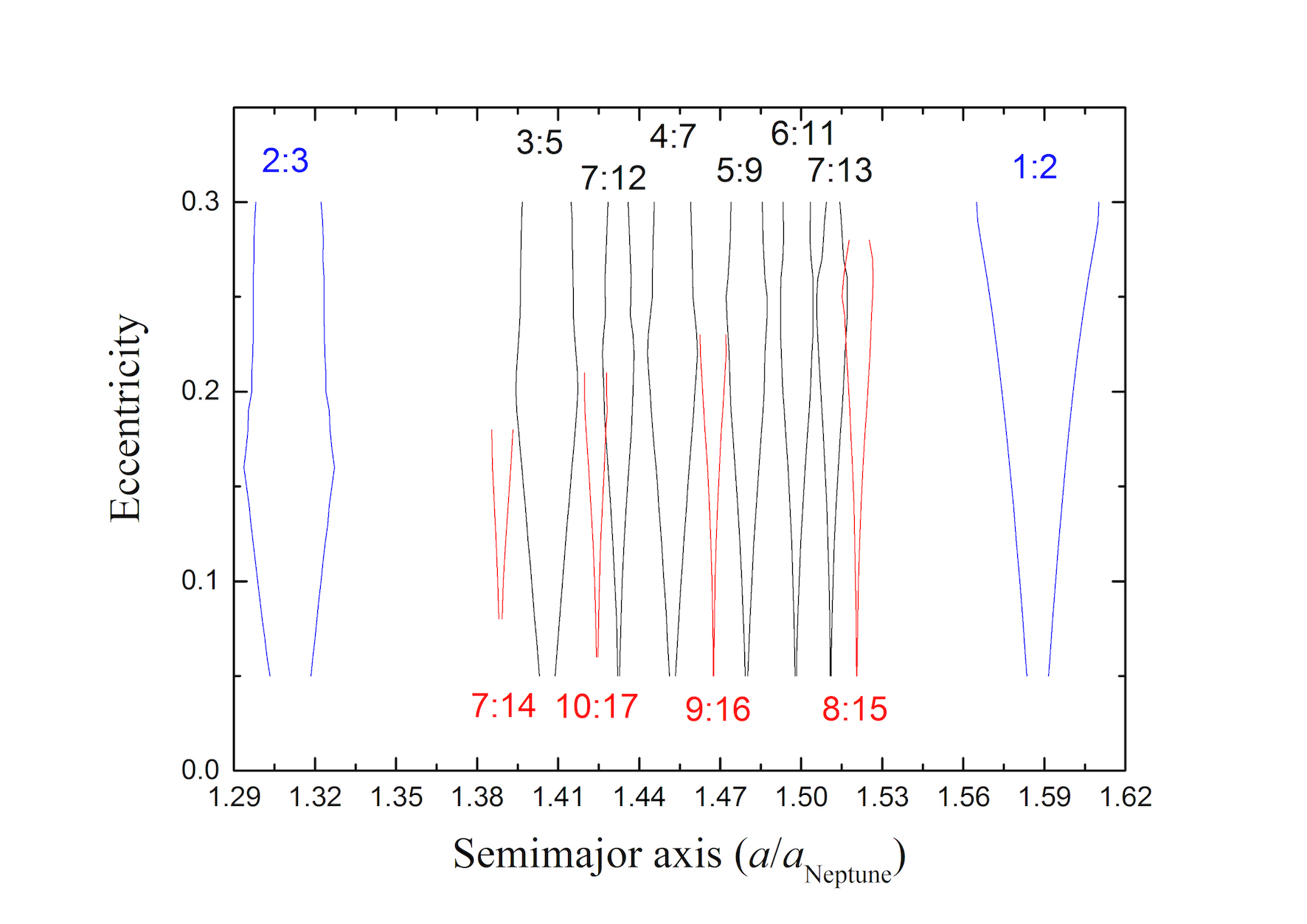}
  \caption{Locations and widths of Neptune's MMRs in the Kuiper belt, according to the phase space structures shown in Fig. \ref{Pspace}. For each MMR, the resonance width is defined as the difference between the maximum and minimum semimajor axes. The 2nd- to 6th-order resonances shown here (black curves) are occupied by the real KBOs. The four 7th-order resonances (red curves) could overlap with lower order resonances nearby, leading to chaotic diffusion. For reference, the 1st-order 2:3 and 1:2 resonances are also plotted (blue curves).}
  \label{width}
\end{figure}

To study the dynamical behavior of small objects in the high-order MMRs with Neptune, we first construct a resonance atlas for the MCKB. Following several previous works by Malhotra and her colleagues \citep{Malh1996, Malh2017, Malh2018, Malh2019}, we adopt the planar circular restricted 3-body problem (PCR3BP) to locate the resonances and calculate their widths. In the framework of the PCR3BP, a massless KBO ($m=0$) moves under the gravitational influence of the Sun ($m_\odot$) and Neptune ($m_N$) in the same plane, i.e. the two massive primaries exert forces on the KBO, but are not affected by the KBO.  

The Sun and Neptune are assumed to have circular orbits about their common center of mass $O$, so they have a fixed angular velocity $n$. It is customary to choose the synodic coordinate system $(x, y)$ that has its origin located at $O$, and rotates at a constant rate $n$. The total mass $(m_\odot + m_N)$, the distance between the Sun and Neptune ($a_N$), and the value of $n$ are all set to be 1. In this non-dimensional system, the gravitational constant $G=1$, and the Sun and Neptune always lie on the $x$-axis at $(-\mu, 0)$ and $(1-\mu, 0)$, respectively, where the mass ratio $\mu=m_N/(m_\odot + m_N)=5.146\times10^{-5}$.

Let the coordinates of the KBO in the synodic system be $(x, y)$, then the equations of its motion can be written as
\begin{eqnarray}
\ddot{x}-2\dot{y}&=&\frac{\partial U}{\partial x},\nonumber\\
\ddot{y}+2\dot{x}&=&\frac{\partial U}{\partial y},
\label{3b}
\end{eqnarray}
where $U$ is as known as a `pseudo-potential':
\begin{equation}
U=\frac{1}{2}(x^2+y^2)+\frac{1-\mu}{r_\odot}+\frac{\mu}{r_N}+\frac{1}{2}\mu(1-\mu),
\label{Ufunction}
\end{equation}
and $r_\odot$ and $r_N$ are the KBO's distance to the Sun and Neptune, respectively:
\begin{eqnarray}
r_\odot^2&=&(x+\mu)^2+y^2,\nonumber\\
r_N^2&=&(x+\mu-1)^2+y^2.
\label{distance}
\end{eqnarray}
The differential system given in equation (\ref{3b}) has a constant of motion: 
\begin{equation}
C(x, y, \dot{x}, \dot{y})=2U(x, y)-(\dot{x}^2+\dot{y}^2),
\label{jacobi}
\end{equation}
which is known as the Jacobi integral.

In order to determine the width of a particular resonance, we have to integrate a large number of nearby test particles. 
We need to note that, in terms of $a$ and $e$, the Jacobi integral can be expressed as  
\begin{equation}
C(a, e)=\frac{1}{a}+2\sqrt{a(1-e^2)}+\mathcal{O}(\mu).
\label{jacobi2}
\end{equation}
So the approximate $C(a, e)$ is different from the exact $C(x, y, \dot{x}, \dot{y})$ only on the order of $\mu\sim10^{-5}$, and independent on the orbital angles $\varpi$ and $\lambda$. We choose the initial values of $(a, e, \varpi, \lambda)$ for test particles similarly to \citet{Malh2017}, as described below. 

The nominal location $a_{res}$ of the $p$:$q$ $(p<q)$ exterior resonance can be calculated using
\begin{equation}
  a_{res}/a_N=\sqrt[3]{q^2/p^2},
\label{nominal}
\end{equation}
where Neptune's semimajor axis $a_N=1$ in the PCR3BP as we defined above. Accordingly, the initial semimajor axes of the test particles for each $p$:$q$ resonance are set to be a fixed value of $a=a_{res}$. The initial eccentricity $e$ ranges from 0 to 0.3. The particles all start at perihelia, thus they have initial $\varpi=0$. The initial value of $\lambda$ is chosen so that the critical resonant angle
\begin{equation}
\sigma=q\lambda-p\lambda_N+(p-q)\varpi,
\label{ResAng}
\end{equation}
varies from $0^{\circ}$ to $180^{\circ}$, where $\lambda_N$ is the mean longitude of Neptune. Since $a$ is fixed at the value of $a_{res}$, in this phase space, $\sigma$ measures the initial separation of a trajectory from the elliptical equilibrium \citep{Li2022}. In order to avoid confusion, we will refer to the specific $p$:$q$ resonant angle as $\sigma_{p:q}$.

The orbital elements of a trajectory can also be written in the inertial frame $(X, Y)$. In the PCR3BP, it is natural to place Neptune on the positive $X$-axis at the time of $t=0$, so the initial value of $\lambda_N$ is $0^{\circ}$. Then the initial $\lambda$ for a particle can be determined using equation (\ref{ResAng}). For each particle, the set of $(a, e, \varpi, \lambda)$ corresponds to the position $(X, Y)$ and velocity $(\dot{X}, \dot{Y})$. Now we can have the initial conditions of a test particle in the synodic system $(x, y)$, as
\begin{eqnarray}
x(0)&=&X(0),\nonumber\\
y(0)&=&Y(0),\nonumber\\
\dot{x}(0)&=&\dot{X}(0)-Y(0),\nonumber\\
\dot{y}(0)&=&\dot{Y}(0)-X(0).
\label{initial}
\end{eqnarray}
where the bracketed digit 0 indicates the value of each coordinate at the time of $t=0$.

Starting from a given set of $(x(0), y(0), \dot{x}(0), \dot{y}(0))$, the motion of the test particle is calculated by integrating equation (\ref{3b}). The duration of the integration is adopted to be 5000 Neptune orbital periods, equal to about $8\times10^5$ yr. Then we can track the time evolution of the critical resonant angle $\sigma$ through a series of transformations, i.e. from the synodic coordinates $(x(t), y(t), \dot{x}(t), \dot{y}(t))$ to the inertial coordinates $(X(t), Y(t), \dot{X}(t), \dot{Y}(t))$, and then to the orbital elements. Neptune's instantaneous $\lambda_N(t)$ can simply be calculated by $t/2\pi$, because the unit of time is $2\pi$ in our non-dimensional system. 

In order to provide a resonant phase space where the libration motion can be clearly visualized, a new critical resonant angle $\phi=\sigma/p$ is introduced, as performed in \citet{Malh2017} and  \citet{Lei2020}. In the Hamiltonian model of the MMRs from our work \citep{Lei2020}, the angle $\phi$ is conjugated to the action $(p/q)\sqrt{\mu(a)}$, where the semimajor axis $a$ is in the unit of $a_N$; while for the other degree of freedom, we use canonical angle $\lambda_N-\varpi$. Then we plot the Poincar\'e surfaces of section $(\phi, a)$, defined by $\lambda_N-\varpi=0$, i.e, one point is recorded each time Neptune completes one revolution. As shown in Fig. \ref{Pspace}, for the 3:5 (left column) and 4:7 (right column) resonances, there are $\mbox{Max}[p, q]=5$ and 7 resonant zones, respectively. We note that in \citet{Malh2019}, the same resonant angle $\phi$ is adopted, but a plot of $(\phi, a)$ on the Poincar\'e sections is recorded at every perihelion passage of the particle. Their results show that, for the $p:q$ exterior resonance, there are $\mbox{Min}[p,q]$ resonance zones in the phase space. So the Poincar\'e sections presented here could reveal completely unfolded phase spaces and provide more detailed dynamical structures for the exterior MMRs \citep[also see][]{Lei2020}.

According to the boundaries of the libration zones in the Poincar\'e section, we can determine the minimum ($a_{\min}$) and maximum ($a_{\max}$) semimajor axes for each MMR, with resonance width defined as $a_{\max}-a_{\min}$. Then we consider a number of Neptune's exterior $p$:$q$ resonances with increasing order ($=q-p>0$) in the MCKB. Fig. \ref{width} presents the resulting resonant widths in the $(a, e)$ plane for these resonances, up through the 7th-order. An obvious feature of the resonant widths is that they become narrower as the order of the resonance increases, and thus the number of RKBOs in a resonance should be fewer as resonance order increases. Indeed, such a trend is visible among the currently observed RKBOs, as shown in Fig. \ref{real}: many are known in the 3:5 and 4:7 resonances, while there is only one known in the 7:13 resonance (6th-order). In Fig. \ref{width} we also notice that the 7th-order resonances (red curves) could overlap with the adjacent lower order resonances. Any objects in these resonances could  escape due to the chaotic diffusion. In the MCKB, no 7th- or higher-order RKBOs have yet been observed with certainty.

As a result, we will only consider the 2nd- to 6th-order resonances within the MCKB for the remainder of this paper, indicated by the black curves in Fig. \ref{width}. We remind the reader that here the planar model is used to calculate the resonant width, while the resonance width generally decreases as the resonant particle's inclination increases \citep[MN2020;][]{Gall2020, namo2020}. Due to this effect, we believe resonance overlap should not be important among the high-order resonances that we consider here.





\section{Distribution of high-order resonators}

\begin{figure}
 \hspace{0cm}
  \centering
  \includegraphics[width=9cm]{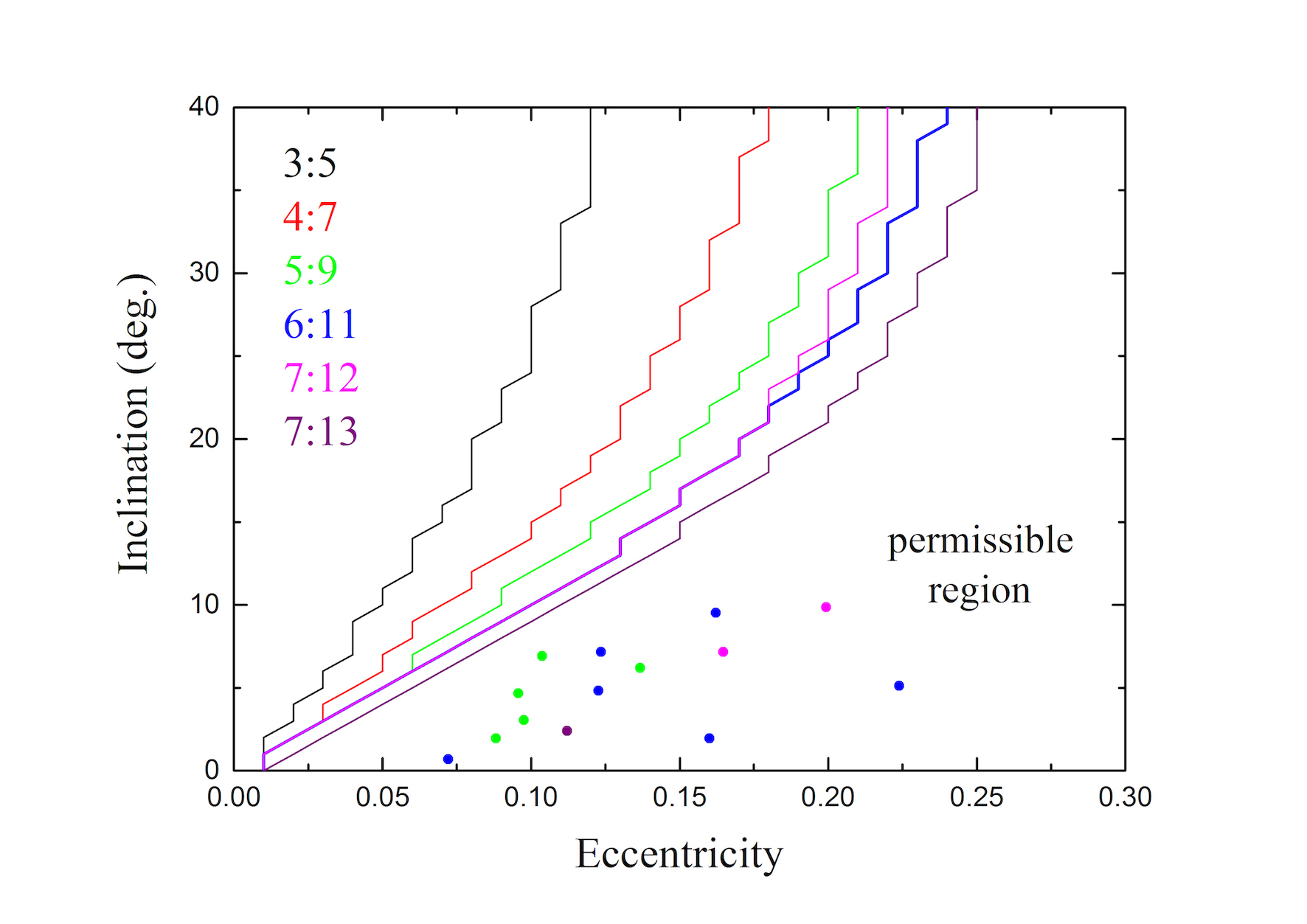}
  \caption{Limiting curves on the $(e, i)$ plane for the high-order resonances in the region of the MCKB, indicated by different colours labelled in the upper left corner. For the highest order 5:9, 6:11, 7:12 and 7:13 resonances, the currently observed RKBOs are also plotted (dots; same colours as limiting curves), and each resonant population lies to the right of the associated limiting curve, i.e. residing in the permissible region.}
  \label{curve}
\end{figure}

It is now well accepted that the RKBOs are objects that formed closer to the Sun, then they were captured by Neptune's MMRs and pushed farther into the Kuiper belt \citep{Malh1993, Malh1995, hahn2005}. According to the capture and migration simulations performed in MN2020, the 4:7 resonant objects must have eccentricities $e\ge e^{4:7}_c(i)$ when they arrived at their current locations, and the critical eccentricity $e_c$ monotonically increases as a function of the resonator's inclination $i$. In other words, the simulated 4:7 resonators reside in the permissible region on the right side of the limiting curve $e^{4:7}_c(i)$, in agreement with the measured $e$ and $i$ distribution for the observed 4:7 RKBOs. A similar evolutionary pathway also drives the observed and simulated 3:5 RKBOs to follow a particular $e$-$i$ distribution, which is constrained by the associated limiting curve $e^{3:5}_c(i)$, as shown in Fig. \ref{simulated35}. 

\begin{table*}
\centering
\begin{minipage}{16cm}
\caption{Statistics of resonant survivors in the $p$:$q$ resonance at the end of the 4 Gyr simulations. The second column ($\Delta_{p:q}$) indicates the maximum widths of resonances which are read from Fig. \ref{width} and then multiplied by $a_N=30.1$ AU. The third column ($N^i_{p:q}$) is the number of captured resonators initially residing in the permissible region of each $p$:$q$ resonance, and we assume $N^i_{p:q}\propto\Delta_{p:q}$. The fourth column ($N^f_{p:q}$) gives the number of resonant survivors, i.e. the objects remaining in the $p$:$q$ resonance after 4 Gyr of orbital evolution, and the corresponding survivability fraction $R_{p:q}$ is presented in the fifth column. Since $N^f_{p:q}$ is very small for the 5:9 and higher order resonances, we carried out additional runs to refine the survivability, to be $\tilde{R}_{p:q}$, as given in the last column.}   
\label{capture}
\begin{tabular}{c c c c c c }        
\hline                 

     Resonance      &   Maximum width        &        Captured resonators     &         Resonant survivors        &            Fraction  (\%)                       &     Refined  fraction  (\%)        \\
           $p$:$q$           &       $\Delta_{p:q}$ (AU)       &                 $N^i_{p:q}$              &              $N^f_{p:q}$               &      $R_{p:q}=N^f_{p:q}/N^i_{p:q}$        &           $\tilde{R}_{p:q}$                  \\
 
\hline

            4:7               &                   0.559                       &                      1050                       &                    420                        &               40                                          &          --    \\
    
            3:5              &                    0.691                       &                      1299                       &                    935                        &               72                                          &          --    \\
                    
            5:9              &                    0.458                       &                        861                        &                    24                         &               2.8                                         &         2.2     \\     
            
            6:11            &                    0.362                       &                        678                        &                   10                          &               1.5                                         &          0.98    \\  
   
            7:12            &                    0.352                       &                        663                        &                     2                          &                0.30                                      &          0.30    \\  
             
            7:13            &                     0.341                      &                        639                         &                    3                          &                0.47                                      &          0.37    \\  

\hline
\end{tabular}
\end{minipage}
\end{table*}

The intrinsic dynamical mechanism accounting for such $e$-$i$ distribution is that, the critical argument $\sigma$ of the 3:5 or 4:7 resonance at high $i$ is allowed to librate only if the condition $e\ge e_c$ is fulfilled. We then suppose that this mechanism should also be valid for other resonances with even higher orders. Accordingly, in Fig. \ref{curve} we plot the limiting curves $e^{p:q}_c(i)$ on the $(e, i)$ plane for all the considered $p:q$ resonances, from the 2nd- to 6th-order. It can be seen that, in addition to the 3:5 and 4:7 resonators discussed before, every observed resonant population we consider follows our prediction and  resides in its permissible region on the right side of the associated limiting curve. 

There are two points that we wish to highlight before proceeding further: (1) in this paper we only consider the $e$-type resonance, where the resonant angle $\sigma$ contains the mean longitudes $\lambda$ and $\lambda_N$, and the particle's longitude of perihelion $\varpi$ (see equation~(\ref{angle35}) for an example). We do not consider the $i$-type or mixed-$(e,i)$-type resonances, where the resonant angle contains the particle's longitude of the ascending node $\Omega$, even though these resonances can occur among the high-inclination KBOs. The odds of detecting these resonators is extremely low; only one object (2014 TZ85) is known among the observed 4:7 RKBOs (see MN2020), and none belongs to the observed 3:5 RKBOs. (2) The limiting curve is calculated only by taking into account the MMR. If the Kozai mechanism is also involved \citep{kozai62, kozai85}, theoretically the resonators could exist in the ($e$, $i$) space outside the permissible region, but the probability is very low. Indeed, we found in the planet migration simulations that all the objects captured into the 3:5 and 4:7 resonances do fall into their respective permissible regions (see Fig. \ref{simulated35} in this paper and Fig. 9 in MN2020).

Based on what we have learned, at the end of planet migration simulations, the captured high-order resonators would not be uniformly distributed in the $(e, i)$ space, but should only occupy the permissible regions. This distribution has to be taken into account for the later long-term evolution, since it places a new constraint on the initial $e$ and $i$ values of the primordial $p:q$ resonators. All of these resonators are required to start from the right side of the associated limiting curve, having $e\ge e^{p:q}_c(i)$. Using such initial conditions, we now construct numerical simulations to investigate the long-term stability of the 3:5, 4:7, 5:9, 6:11, 7:12 and 7:13 resonances. A systematic analysis of these six resonances at large inclinations could provide us interesting information not only on the possible distribution of the high-order RKBOs, but also on the dynamical structure of the entire MCKB.

\subsection{Initial conditions and pre-runs}

First, we start by generating test resonators for the 4:7 MMR. Particles are assumed to have initial inclinations and eccentricities randomly distributed in the range of $i_0=0^{\circ}-40^{\circ}$ and $e_0=0-0.3$. Both of these ranges are chosen to agree with the range of currently observed objects in the MCKB. Next we calculate the critical eccentricity $e^{4:7}_c$ at $i_0$ for each particle, and we keep it as a test resonator if its $e_0$ is equal or greater than $e^{4:7}_c(i_0)$. In this way, we introduce tens of thousands test resonators with reasonable $(e, i)$ pairs for which the libration of $\sigma_{4:7}$ is allowed. The test resonators' initial semimajor axes are set to be $a_{res}$, i.e. at the nominal location of the 4:7 resonance defined in equation (\ref{nominal}); and the three orbital angles are chosen such that all of these resonators begin with $\sigma_{4:7}=180^{\circ}$, $\omega=90^{\circ}$ and $\Omega=0^{\circ}$, as we did in MN2020.

Second, in order to mimic the objects captured into the 4:7 MMR during Neptune's migration, we need to check whether the test resonators can exhibit libration of $\sigma_{4:7}$. For this purpose, pre-runs are carried out in the framework of the present outer Solar system, by employing the SWIFT\_RMVS3 symplectic integrator with a time step of 0.5 yr \citep{Levi1994}. We numerically calculated the resonant amplitudes $A_{\sigma_{4:7}}$ of test resonators, by measuring the maximum deviations of $\sigma_{4:7}$ from the libration center at $180^{\circ}$ over the 10 Myr integration. Out of 3000 test resonators, $N^i_{4:7}=1050$ objects fulfilling the criteria $A_{\sigma_{4:7}}<175^{\circ}$ were selected for the later long-term runs, and they are regarded as `captured resonators' hereafter. Different from the discrete $e_0$ and $i_0$ values adopted in MN2020, these captured resonators cover the entire permissible $(e, i)$ region of the 4:7 MMR, lying to the right of the red limiting curve plotted in Fig. \ref{curve}.

Repeating the above two-step procedure (i.e. generating test resonators in the permissible region, and then selecting captured resonators via the 10 Myr short-term runs), we also produce initial captured resonator populations for the 3:5, 5:9, 6:11, 7:12 and 7:13 resonances. For each of these high-order resonances, we determine the number $N^i_{p:q}$ of captured resonators using the ratio of each resonance's maximum width to that of the 4:7 resonance, as
\begin{equation}
  N^i_{p:q}=\frac{\Delta_{p:q}}{\Delta_{4:7}}N^i_{4:7},
\label{number}
\end{equation}
where $\Delta_{p:q}$ is the maximum width of the $p:q$ resonance, and $N^i_{4:7}(=1050)$ is the number of the 4:7 initial captured resonators as we set above. This is a reasonable assumption since the width of a resonance is positively correlated with the strength of the resonance \citep{Gall2019}. The stronger the resonance is, theoretically the more efficient it is at trapping and retaining particles while sweeping outwards, leading to more captured resonators. We note that, as shown in Fig. \ref{width}, the resonance width in semimajor axis actually depends on the eccentricity. For scaling the resonance's relative strength, it may be more precise to utilize the area of the resonance zone. However, the maximum width serves as a good approximation and it can be directly read from Figure \ref{width}. The measured maximum widths $\Delta_{p:q}$ of six considered resonances are listed in Table \ref{capture}. For ease of comparison, the unit of $\Delta_{p:q}$ is transformed to AU by multiplying by Neptune's semimajor axis of $a_N=30.1$ AU. To simply validate the adoption of the resonance's maximum width in Equation (\ref{number}), let us consider the 4:7 and 3:5 resonances. According to Table 1 in \citet{Malh2019}, the area ratio of their resonant zones is about 1.32, while the ratio of their maximum widths is ${\Delta_{3:5}}/{\Delta_{4:7}}=1.24$.  These two ratios are close enough to each other that we feel validated in using the simpler maximum width for our population ratios. Now according to equation (\ref{number}), the number $N^i_{p:q}$ of captured resonators can be calculated for a particular $p:q$  resonance, as presented in the third column of Table \ref{capture}. We note that the total number of test resonators (e.g. 3000 for the 4:7 resonance) could vary for different resonances, as long as it is large enough to produce the required number $N^i_{p:q}$ of captured resonators.


Although the evaluation of $N^i_{p:q}$ with respect to a given $N^i_{4:7}$ is a reasonable starting point, there are other factors within resonance sweeping and capture models that could affect these relative numbers. The number and orbital distributions of disk particles should be about the same when each individual resonance sweeps through, as assumed in \citet{morb05}. Additionally, when these resonances passes through a particular location, they should have similar migration rates. Although these two factors are ignored in our calculations, they could approximately cancel each other out. Let us take the 4:7 and 3:5 resonances for an illustration: as Neptune moves outwards, the 4:7 resonance sweeps ahead of the 3:5 resonance through the outer planetesimal disk, capturing some of the planetesimals and thus leaving fewer planetesimals for the 3:5 resonance that follows. On the other hand, when the 3:5 resonance sweeps through the same radial location in the planetesimal disk, it migrates at a lower speed than the preceding 4:7 resonance (see Appendix B for details). Thus the slower pace of the 3:5 resonance would lead to the higher capture efficiency of planetesimals \citep{Meli2000, Li2006}, even though fewer are left behind to capture.

There are certainly ways to increase the accuracy of sweeping migration capture simulations, 
such as accounting for the factors mentioned above, and including stochastic jumps in Neptune's migration due to the scattering of discrete planetesimals \citep{zhou2002, nesv16}. 
But our methods here are satisfactory for placing some constraints on various models of the Kuiper belt formation, by simulating the long-term orbital evolution of the captured resonators from our pre-runs. We will evaluate the population ratios and orbital distributions of the surviving populations of captured resonators from our simulations, and then compare them to the observations. And, particular attention is also paid to those simulated resonators that experienced chaotic diffusion in the high-order resonances, and became non-resonant KBOs with very different final $e$ and $i$.

\subsection{Long-term runs}

For each of the six considered high-order resonances, we numerically integrate the captured resonators from the pre-runs for the age of the Solar system (4 Gyr). After the long-term simulations have been accomplished, we continue the integrations for another 10 Myr to classify the behaviors of the surviving objects. We also calculate these objects' average semimajor axes, eccentricities and inclinations, which are recorded as the final orbital elements $a_f$, $e_f$ and $i_f$, respectively. Among the $N^i_{p:q}$ intitial captured resonators, the surviving objects are referred to as `resonant survivors' if they librate in their individual resonances during the 10 Myr classification simulations, while the other particles that left the resonances but can remain in the MCKB are called `non-resonant survivors'. The final orbital distributions of these two simulated populations are presented in Fig. \ref{Res-NonRes}, with further discussion presented in Section~\ref{sec:nonres}.

\subsubsection{Resonant survivors}

\begin{figure*}
  \centering
  \begin{minipage}[c]{1\textwidth}
  \vspace{0 cm}
  \includegraphics[width=9cm]{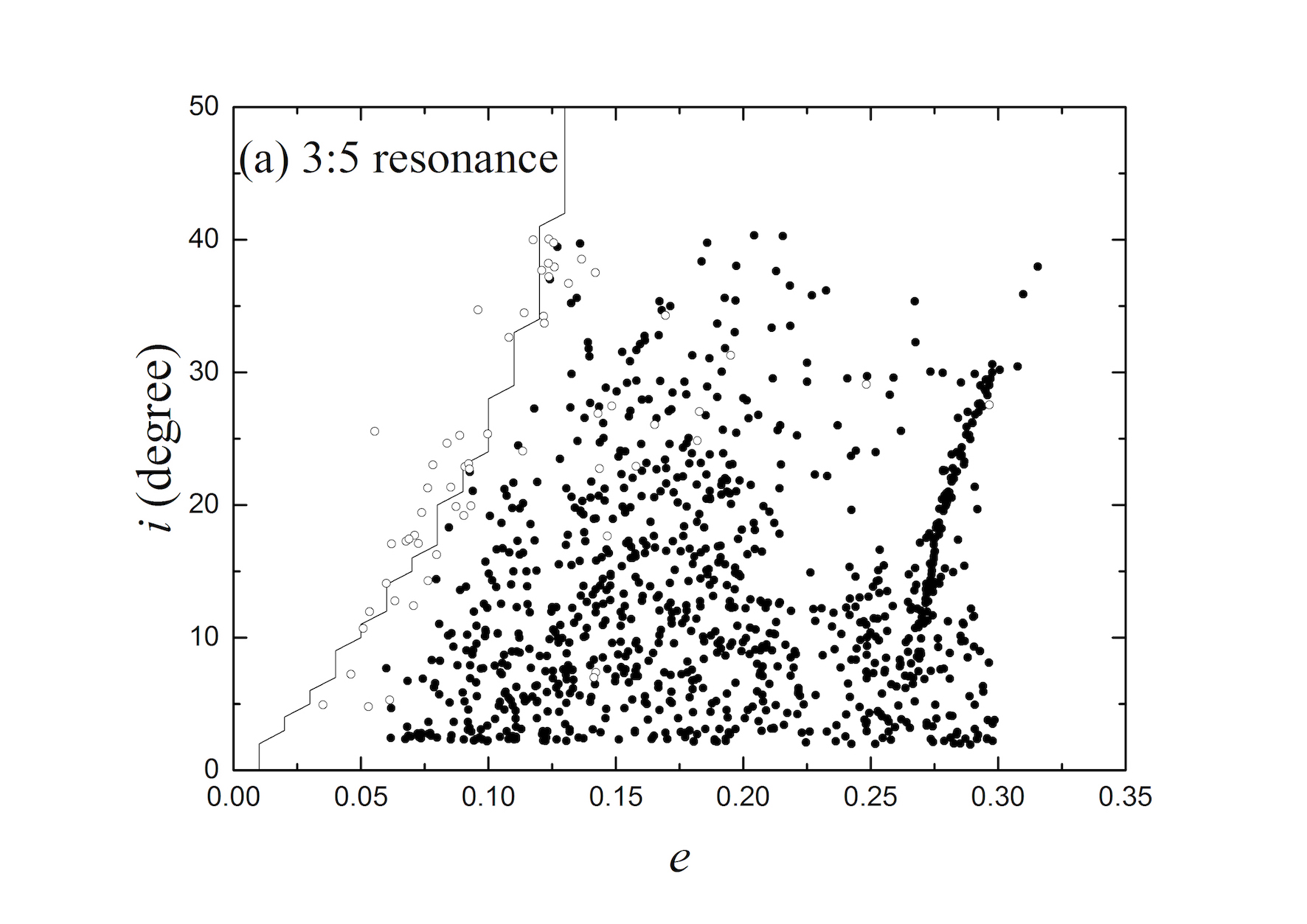}
  \includegraphics[width=9cm]{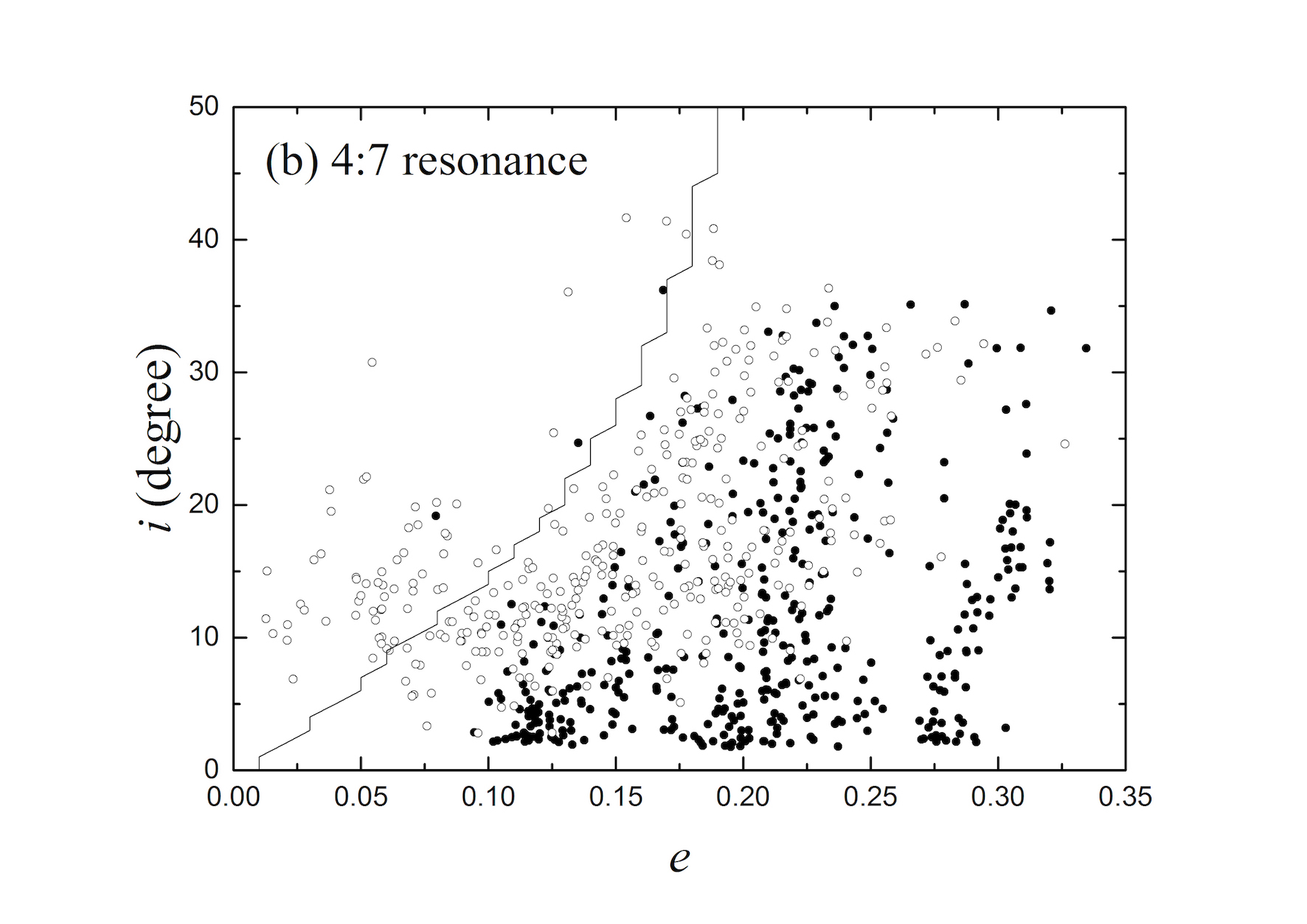}
  \end{minipage}
  \begin{minipage}[c]{1\textwidth}
  \vspace{0 cm}
  \includegraphics[width=9cm]{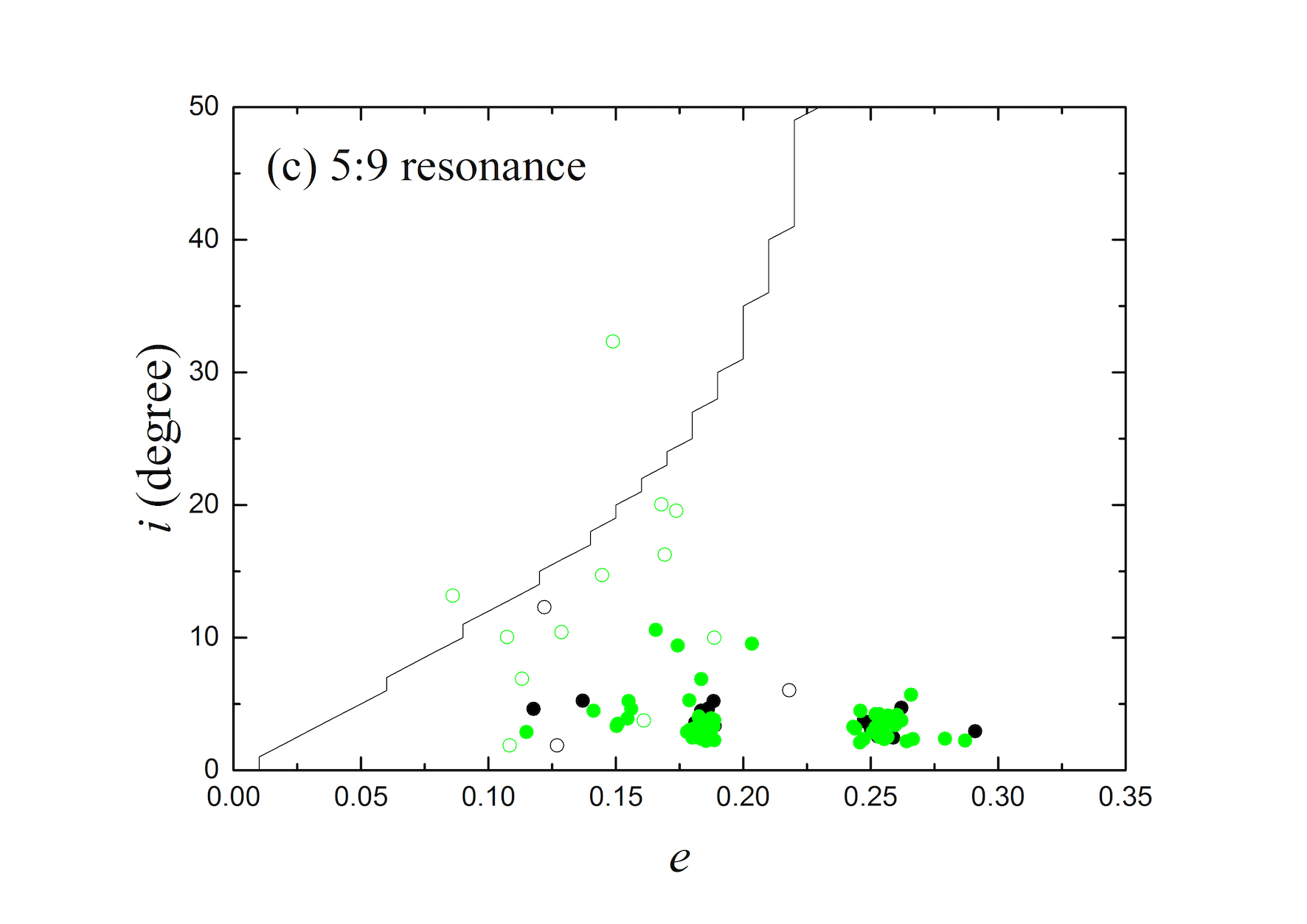}
  \includegraphics[width=9cm]{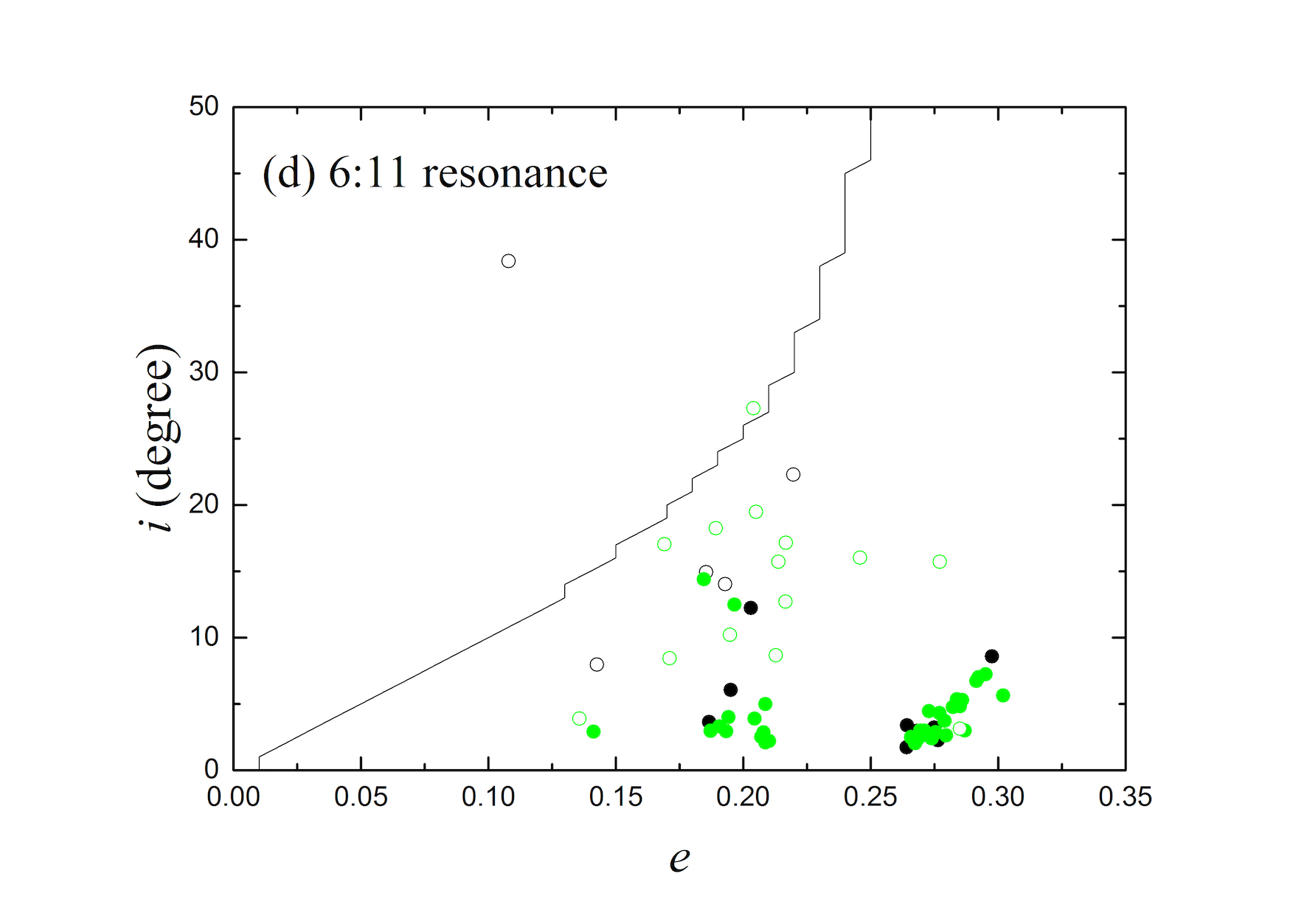}
  \end{minipage}
  \begin{minipage}[c]{1\textwidth}
  \vspace{0 cm}
  \includegraphics[width=9cm]{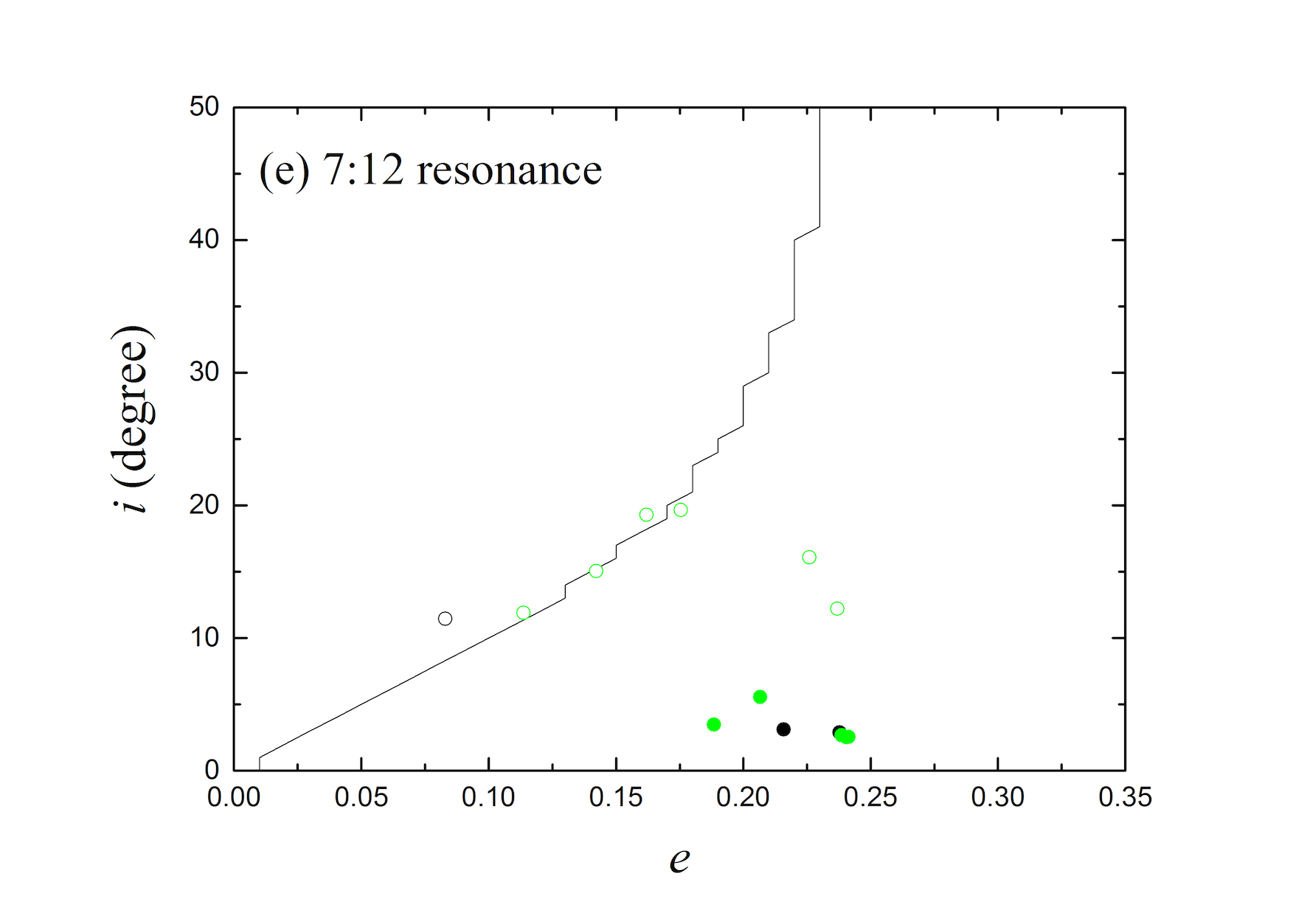}
  \includegraphics[width=9cm]{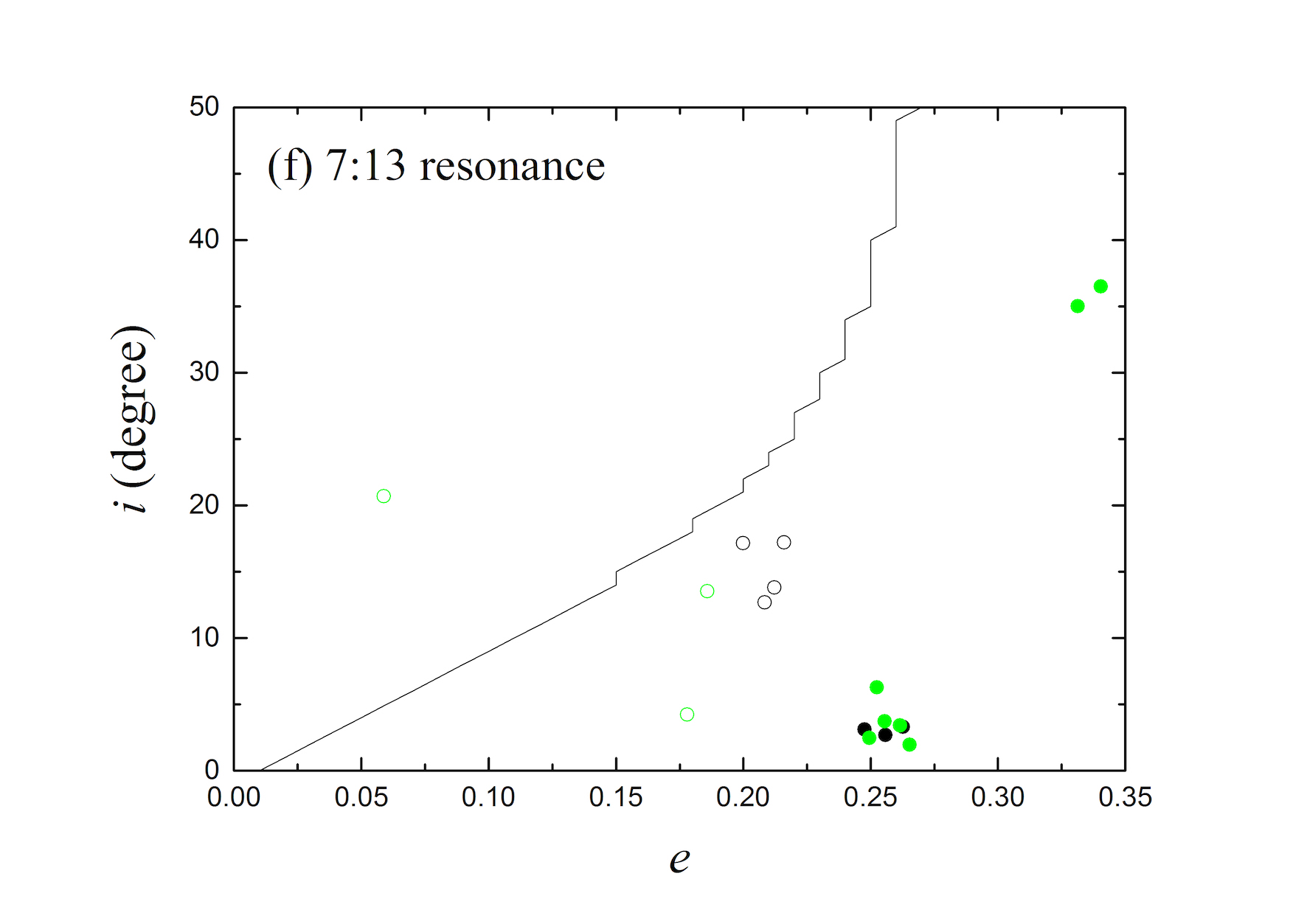}
  \end{minipage}
   \vspace{0 cm}
  \caption{Final eccentricities and inclinations of surviving particles from the 4 Gyr evolution simulations, for each of the six high-order resonances that we considered. The filled and empty circles represent the resonant and non-resonant survivors, respectively. Notice that for each resonance, nearly all the resonant survivors (filled circles) are inside the permissible region to the right of the limiting curve (black line). Some of the non-resonant survivors (empty circles) have crossed the limiting curve and end up on high-inclination and small-eccentricity orbits, even though they originally started within the permissible region. The green symbols indicate the objects from the extra runs for the 5:9 and higher order resonances, in order to increase statistical significance.}
 \label{Res-NonRes}
\end{figure*}  

The resonant survivors from the 4 Gyr simulations are indicated by the black dots in Fig. \ref{Res-NonRes}. We first focus on the two strongest MMRs in the MCKB, i.e. the 3:5 and 4:7 resonances, as shown in the upper two panels. The key result is that a large fraction of the 3:5 and 4:7 resonators can remain in the individual resonances for the full 4~Gyr since they were captured, and these objects consistently occupy the full range of their respective permissible regions to the right of the associated limiting curves (black lines).  We do observe that three of the hundreds of 4:7 resonant survivors settle outside the permissible region. This small fraction of exceptions is not unexpected, due to the occasional Kozai-affected particle as we noted in MN2020. No exceptions are observed in the 3:5 resonant survivors; all particles stay within the permissible region after 4 Gyr of integration.

To quantitatively describe the long-term stability of the MMRs in the MCKB, the statistics of individual resonances are shown in Table \ref{capture}. Among the 1050 captured 4:7 resonators from the pre-run, a total of $N^f_{4:7}=420$ objects remain in libration after the 4 Gyr evolution.  In other words, the long-term survivability of the 4:7 resonators, $R_{4:7}$, is about 40 per cent. The survivability is even higher for the 3:5 resonators, at $R_{3:5}=72$ per cent. As represented by filled circles in Figs. \ref{Res-NonRes}(a) and (b), at the end of the integration the 3:5 and 4:7 simulated resonators nearly fulfill their respective permissible $(e, i)$ regions, and thus keep the original $e$-$i$ distributions in general. This shows that the 4 Gyr evolution after migration does not have a big effect on the overall distribution of these two resonant populations. One may additionally notice that, among either resonant population, there are a bunch of objects that are clumped together in a line at $e=0.25$-0.3 and $i=10^{\circ}$-$30^{\circ}$. We checked with our simulation data and found such clumps are associated to the Kozai mechanism due to the objects' quite large eccentricities and inclinations.

However, for the 5:9 resonance, our simulations show that very few objects maintain resonant behaviour at the end of the long-term integration, with a rather low survivability of $R_{5:9}<3$ per cent. The survivability fraction decreases even lower to only $\sim1$ per cent for the higher order 6:11, 7:12 and 7:13 resonances. Because of the very small numbers of resonant survivors for these four resonances (see $N^f_{p:q}$ in Table \ref{capture}), we conduct additional runs to enlarge their sample sizes. For each of these highest order resonances, we generate two new sets of captured resonators, each of which contains the same number $N^i_{p:q}$ of particles as given in Table \ref{capture}. Then the systems are integrated for 4 Gyr. Together with the previous sample set, we have now increased the total number of captured resonators to $3\times N^i_{p:q}$, and accordingly we calculate the refined survivability $\tilde{R}_{p:q}$, as listed in the last column of Table \ref{capture}. We can see that there is no apparent change in the value of $R_{p:q}$ with a larger integrated sample size. Thus, the extremely low survivability of the 5:9, 6:11, 7:12 and 7:13 resonators is confidential but not just an effect of small number statistics. In Figs. \ref{Res-NonRes}(c)--(d), the resonant survivors from additional simulations are indicated by the green symbols.

The individual fractions of resonant survivors listed in Table \ref{capture} imply that the numbers of simulated objects in the 5:9,  6:11, 7:12 and 7:13 resonances should be of roughly the same order of magnitude, and at least an order of magnitude less than the 3:5 or 4:7 resonances. We feel this result is, in general, compatible with the number ratios among the observed RKBO populations as shown in Fig. \ref{real}; and we note that the small range in semimajor axes and similar expected orbital distributions mean there are not likely to be significant differences in the observational biases for these particular resonances. Additionally, Fig. \ref{Res-NonRes} shows that nearly all of the simulated resonators survive in their respective permissible regions, also consistent with the current observations (see Fig. 4(a) in MN2020, Figs. \ref{simulated35} and \ref{curve}). Thus this paper extends our previous limiting curve theory developed in MN2020 for the 4:7 resonance to all the high-order resonances (up to 6th-order) in the MCKB region. Another point to highlight is that, as shown in Figs. \ref{Res-NonRes}(c)--(f), the overwhelming majority of the 5:9, 6:11, 7:12 and 7:13 simulated resonators are found on lower inclination orbits with $i<10^{\circ}$, similar to the inclination range of the corresponding real RKBOs (see the orange, purple, red and blue dots in Fig. \ref{real}). Our simple model of capture and in-situ evolving seems to have correctly reproduced some intrinsic structures of the RKBOs in the MCKB region.


\subsubsection{Non-resonant survivors} \label{sec:nonres}

In our simulations, we find that some initial captured resonators escape from MMRs, but they remain in the MCKB region at the end of the long-term integration, as indicated by the empty circles in Fig. \ref{Res-NonRes}. These particles had been inside high-order resonances for some length of time, where chaotic motion could cause irregular variations of their $e$ and $i$ \citep{Lyka2005b}, eventually pushing them outside the resonance space. A subset of escapees crossed the limiting curve and settled on the left side, and such objects seem to be quite plentiful in Fig. \ref{Res-NonRes}(b) associated to the 4:7 resonance.  Thus, a testable prediction of our model is that this mechanism should generate a number of classical KBOs (i.e. outside the MMRs) moving on highly inclined but nearly circular orbits, which is a population that is very difficult to produce by other known emplacement mechanisms. 

\begin{figure}
 \hspace{0cm}
  \centering
  \includegraphics[width=9cm]{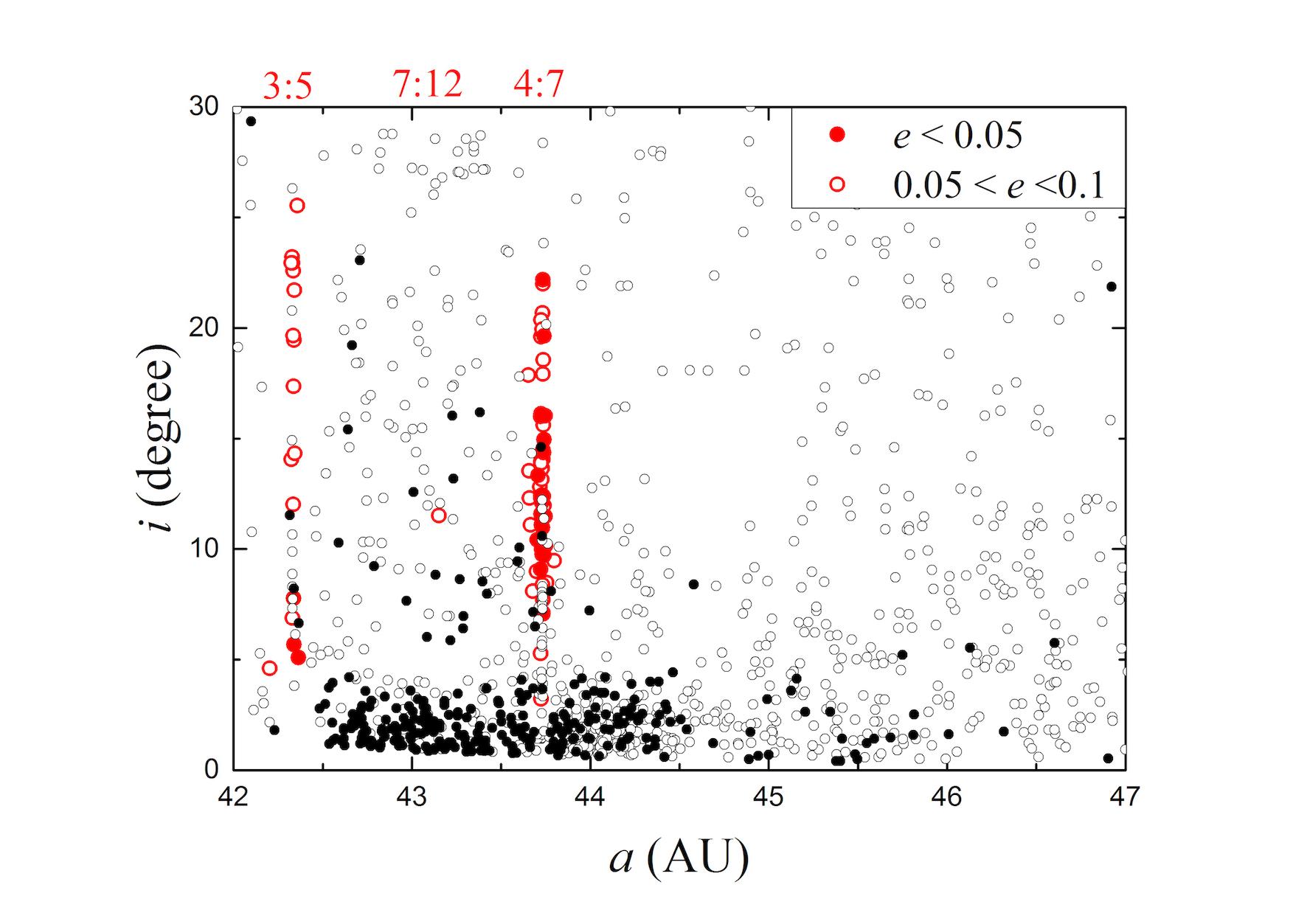}
  \caption{Distribution of semimajor axes and inclinations for the non-resonant objects with small $e < 0.05$ (filled circles) and $0.05 < e < 0.1$ (empty circles) in the MCKB. The red symbols indicate the simulated objects that escape from the 3:5, 7:12 and 4:7 resonances (left to right), from our 4 Gyr long-term runs. The black symbols indicate the real observed KBOs.}
 \label{NonRes}
\end{figure}

In Fig. \ref{NonRes},  the red symbols show the $a$-$i$ distribution for all the non-resonant survivors with $e<0.1$ that leaked from the six simulated resonances. For comparison, the distribution of the real observed classical KBOs with $e<0.1$ has also been presented by the black symbols in this figure. We can see that, for both the simulated and observed non-resonators, the objects with eccentricities below 0.05 mainly concentrate in the inner part of the MCKB, i.e. having $a<45$ AU. This peculiar eccentricity distribution was noticed in the low-inclination population and reproduced by having Neptune's outward migration start from an eccentric orbit that circularizes subsequently \citep{Morb2014}. Here, we observe a similar but weaker trend for the high-inclination population. However, we do not intend to concentrate on the emplacement of the smallest $e$ population, but will aim at the overall distribution of all the non-resonant survivors with $e<0.1$ in the MCKB.

Our work is focused on the high-inclination KBOs with $i>10^{\circ}$, which are believed to have been formed closer to the Sun and suffered close encounters with one or more planets \citep{Gome2003, nesv15, ribe19, Gome2021}. During these encounters, both the inclinations and eccentricities of these objects would be pumped up, leading to a deficit of nearly-circular objects with high-$i$ in the MCKB \citep{Levi2008}. Using our limiting curve theory for the RKBOs, we also expect correlation between the inclination and eccentricity, as the original particles (i.e. captured resonators) possessing higher inclinations tend to occupy more eccentric orbits. According to the geometries of the individual permissible regions sketched in Fig. \ref{curve}, initially the high-$i$ ($>10^{\circ}$) and low-$e$ ($<0.1$) test resonators generated in the pre-runs should be quite rare, but they become more numerous through the mechanism of chaotic evolution and resonance escape during the 4 Gyr evolution as we described above. Fig. \ref{NonRes} shows that (see red symbols), within such high-$i$ and low-$e$ ranges, our simulations produce a number of non-resonant survivors.  

We find that the simulated non-resonant survivors with high-$i$ and low-$e$ are concentrated very close to several resonances, so some other mechanism is needed to make them radially disperse.  The redistribution of their semimajor axes could possibly occur as Neptune and its MMRs migrated outwards. \citet{kaib16} proposed that Neptune took about 300 Myr to reach its current location, in order to reproduce a measurable population of high-perihelion objects residing near the 1:3 MMR. This slow migration scenario, which results in high-$i$, low-$e$ non-resonant survivors concentrated close to resonances, was supported by the distribution of detected KBOs in and near the 2:5 and 1:3 MMRs in the OSSOS observations \citep{lawl19}. \citet{kaib16} showed that a 300 Myr evolution timescale of Neptune's grainy migration is long enough for Kozai cycling within resonances combined with chaotic motion to induce a pronounced decrease in the typical eccentricities for objects that fall out of resonances during migration. The typically high-perihelion orbits for these resonant dropouts are then protected from  close encounters with Neptune. This process is similar to the chaotic evolution and resonance escape mechanism as we noted above. Thus it is quite possible that the slow migration of Neptune may generate high-$i$ and low-$e$ non-resonators that widely dispersed in the MCKB. The population of non-resonant survivors is a byproduct of this work, and a detailed exploration would be carried out in the future.


\section{Conclusions and discussion}

In our previous work about the dynamics of the 3rd-order 4:7 MMR with Neptune (MN2020), we found that the libration zone is constrained by a limiting curve $e_c(i)$, where $e_c$ is the critical eccentricity and increases with the inclination $i$. Only if the condition $e\ge e_c$ is satisfied, the resonant angle is permitted to librate. This implies that, on the $(e, i)$ plane, the distribution of the 4:7 resonators would not be uniform. Accordingly, we infer that this theory could be generalized to a large variety of high-order (i.e. second or higher orders) MMRs in the MCKB, which we explored in this paper. 


We first consider the 3:5 MMR, which is a 2nd-order resonance where our limiting curve theory should also be applicable. We start with 4:7 MMR migration capture simulation data from MN2020, and found some particles are picked up by the 3:5 MMR following behind the 4:7 MMR. All of the simulated 3:5 resonators at the end of the migration are associated with the eccentricity-type resonance, and also lie in the permissible region where the corresponding resonant angle is allowed to librate. The same features hold for the currently observed real 3:5 RKBOs. These 4:7 and 3:5 simulation results show we only need to consider the eccentricity-type resonance, which seems to be the dominating mode of the high-order MMRs. In further simulations, we determine a set of lower limits to the eccentricity for high-inclination orbits for the 2nd- to 7th-order MMRs in the MCKB, and we provide the most plausible $e$-regions where the inclined RKBOs may inhabit.

In order to meet goal $(\mathbb 1)$ raised in the introduction, revealing the global dynamics of high-order resonances, we first set up the PCR3BP model to investigate the phase space structures for each resonance. By introducing a new resonant angle, we can clearly locate the boundaries of the libration zones, and accordingly calculate the resonance widths. We notice that: (1) the width of the 6th-order 7:13 resonance is quite narrow, this may explain why there is only one 7:13 RKBO observed so far; (2) the 7th- and higher-order resonances could overlap with nearby lower-order resonances, leading to instability due to chaotic diffusion.

Next, as a generalization of our limiting curve theory, we conduct numerical simulations to predict the orbital distribution of the objects in the 3:5, 4:7, 5:9, 6:11 7:12 and 7:13 (eccentricity-type) MMRs with Neptune. By adopting initial $(e, i)$ pairs within each MMR's permissible region, which lies to the right of the limiting curve, a number of initial test resonators are generated. To recreate the objects captured into Neptune's resonances, we select those test resonators that are stably librating for 10 Myr, and we use them as our initial captured resonators for longer simulations. For each $p:q$ resonance, the number $N^i_{p:q}$ of initial captured resonators is assumed to be proportional to the resonance's maximum width. With determined initial conditions of these six high-order resonances, we numerically integrate the orbits of $N^i_{p:q}$ captured resonators for the age of the Solar system (4 Gyr). The final distributions of the surviving resonators show that nearly all of the simulated objects that are still librating are inside their respective permissible regions. This result validates our conjecture that the limiting curve theory developed in MN2020 could be applied to all the high-order resonances in the MCKB for constraining the positions of the RKBOs in the $(e, i)$ space.

To achieve the goal $(\mathbb 2)$ of this paper, which is to estimate the number of objects associated with individual high-order resonances, we analyze the various final states of the initially resonant particles from the long-term simulations performed above. The survival rates of captured resonators in the 3:5 and 4:7 MMRs are quite high, about 72 per cent and 40 per cent, respectively. Surviving particles in these two resonances occupy the entire permissible regions, with inclinations up to $40^{\circ}$. Meanwhile, only about 1 per cent of captured resonators from the weaker 5:9, 6:11, 7:12 and 7:13 MMRs remain in the individual resonances after 4 Gyr of evolution, and nearly all of them are found on lower inclination orbits with $i<10^{\circ}$. Furthermore, by taking into account the initial numbers of captured resonators associated with different resonances, our simulations estimate that: (1) the 3:5 and 4:7 MMRs should contain comparable numbers of RKBOs, with a slightly higher population in the former one; (2) the populations in other higher order (5:9, 6:11 7:12 and 7:13) resonances are about an order of magnitude smaller than the 3:5 and 4:7 resonant populations. These results are in a good agreement with the current observation, and thus can provide a prediction about the intrinsic numbers and orbital distributions of all the RKBOs in the region of the MCKB. We want to emphasize that, even under different planet migration scenarios that predict different initial numbers of captured resonators (as we discussed in Sect. 4), the survival rate $R_{p:q}$ for each resonance obtained from the long-term simulations in this work (see Table \ref{capture}) would remain unchanged.

In addition to the resonant survivors which were the main focus of our work, we also find a population of non-resonant survivors. It is very interesting to notice that before these simulated objects leaked out of their resonances, they had experienced significant changes in $e$ and $i$ due to chaotic motion. After crossing their respective limiting curves, they became non-resonators with highly inclined but nearly circular orbits. This population may help provide a solution to goal $(\mathbb 3)$ raised at the end of the introduction, to explain the deficit of low-$e~(<0.1)$ objects in the MCKB noted in previous works, especially for the high-inclination population. Nevertheless, possible mechanisms of later redistribution of their semimajor axes, to be widely spread in the MCKB, need further detailed investigation.


\section*{Acknowledgments}

This work was supported by the National Natural Science Foundation of China (Nos. 11973027, 11933001, 12150009, 12073011), and National Key R\&D Program of China (2019YFA0706601). The authors would like to express their thanks to the anonymous referee for the valuable comments.

\section*{Data Availability}

The data underlying this article are available in the article and in its online supplementary material.

\clearpage
\onecolumn

\clearpage
\appendix

\section{Resonate KBOs in the MCKB between 42 and 47 AU}
\label{sec:RKBOs list}

\begin{longtable}{l c c c c r r r r r c}
\caption{Orbital elements of the observed resonant KBOs displayed in Fig. \ref{real}, at epoch 1969 May 31. The original data is obtained from the MPC. `Opps' denotes the number of oppositions over which the KBO has been observed, and `Res' indicates the specific MMR in which the KBO resides. The three angles $\Omega$, $\omega$ and $M$ are the longitude of ascending node, argument of perihelion and mean anomaly, respectively. The resonant amplitude $A_{\sigma}$ is determined from a full 10 Myr simulation of each object. For reference, the last column provides the orbit uncertainties characterised by measured semimajor axes $\Delta a/a$, which are as small as $<0.16\%$. Note that only two 6:11 resonators (indicated by a star symbol) exhibit relatively larger $\Delta a/a\sim0.1$, they are considered to be insecure resonators and only listed in this table for reference, but not included in Fig. \ref{real} or the main text.} 
\label{realRKBOlist} \\

\hline \multicolumn{1}{c}{MPC Designation} & \multicolumn{1}{c}{Opps.} & \multicolumn{1}{c}{Res.} & \multicolumn{1}{c}{$a$ (AU)} & \multicolumn{1}{c}{$e$} & \multicolumn{1}{c}{$i$ ($^{\circ}$)} & \multicolumn{1}{c}{$\Omega$ ($^{\circ}$)} & \multicolumn{1}{c}{$\omega$ ($^{\circ}$)} & \multicolumn{1}{c}{$M$ ($^{\circ}$)} & \multicolumn{1}{c}{$A_{\sigma}$ ($^{\circ}$)} & \multicolumn{1}{c}{$\Delta a/a$ (per cent)} \\ \hline 
\endfirsthead

\multicolumn{11}{c}%
{{ \tablename\ \thetable{} -- continued from previous page}} \\
\hline \multicolumn{1}{c}{MPC Designation} & \multicolumn{1}{c}{Opps.} & \multicolumn{1}{c}{Res.} & \multicolumn{1}{c}{$a$ (AU)} & \multicolumn{1}{c}{$e$} & \multicolumn{1}{c}{$i$ ($^{\circ}$)} & \multicolumn{1}{c}{$\Omega$ ($^{\circ}$)} & \multicolumn{1}{c}{$\omega$ ($^{\circ}$)} & \multicolumn{1}{c}{$M$ ($^{\circ}$)} & \multicolumn{1}{c}{$A_{\sigma}$ ($^{\circ}$)} & \multicolumn{1}{c}{$\Delta a/a$ (per cent)} \\ \hline 
\endhead

\hline \multicolumn{11}{c}{Continued on next page} \\ \hline
\endfoot

\hline \hline
\endlastfoot

2015	VK167	&	4	&	3:5	&	42.253	&	0.268	&	12.0 &	220.8	&	118.6 	&	49.7 	&	52.8	& 0.021 \\
2015	VJ167	&	4	&	3:5	&	42.034	&	0.203	&	3.5  &	97.0	&	41.3 	&	301.4	&	74.7	& 0.024 \\
2015	VH167	&	3	&	3:5	&	42.205	&	0.140	&	0.7	 &	66.5 	&	293.6 	&	45.2	&	96.7	& 0.032 \\
2015	VG167	&	3	&	3:5	&	42.142	&	0.250	&	4.8	 &	67.7 	&	69.5 	&	308.3 	&	33.4 	& 0.046 \\
2015	RK278	&	3	&	3:5	&	42.458	&	0.158	&	9.8	 &	351.1	&	305.3 	&	62.6 	&	48.6 	& 0.038 \\
2015	RJ278	&	3	&	3:5	&	42.279	&	0.154	&	5.2	 &	358.3	&	194.6 	&	178.3 	&	149.2	& 0.009 \\
2015	KK174	&	3	&	3:5	&	42.430	&	0.233	&	9.6	 &	111.4 	&	350.1	&	113.1	&	24.8 	& 0.074 \\
2015	GO55	&	3	&	3:5	&	42.516	&	0.126	&	8.4  &	43.8	&	221.5	&	314.6	&	154.3	& 0.023 \\
2014	WM509	&	3	&	3:5	&	42.187	&	0.118	&	24.2 &	88.9 	&	318.3	&	3.6 	&	72.9	& 0.005 \\
2014	TA86	&	7	&	3:5	&	42.227	&	0.108	&	9.3	 &	219.3	&	140.1	&	39.6 	&	81.4 	& 0.003 \\
2014	SJ349	&	7	&	3:5	&	42.147	&	0.162	&	8.9  &	123.8	&	296.3 	&	357.4 	&	40.6	& 0.004 \\
2014	QZ441	&	7	&	3:5	&	42.282	&	0.078	&	6.5	 &	335.0 	&	176.1 	&	241.8 	&	95.6	& 0.005 \\
2014	OL394	&	8	&	3:5	&	42.158	&	0.270	&	4.6	 &	208.4	&	171.3	&	5.4	    &	131.6 	& 0.004 \\
2014	OK394	&	7	&	3:5	&	42.252	&	0.164	&	4.1  &	128.4 	&	247.5 	&	30.9	&	81.5 	& 0.003 \\
2014	KT101	&	5	&	3:5	&	42.505	&	0.251	&	9.9  &	66.4 	&	206.6	&	353.9	&	145.1 	& 0.005 \\
2014	DK143	&	9	&	3:5	&	42.458	&	0.157	&	10.9 &	153.6	&	337.6 	&	90.8 	&	36.0	& 0.004 \\
2013	UT22	&	6	&	3:5	&	42.261	&	0.215	&	29.3 &	194.2	&	137.9	&	54.1	&	70.0	& 0.024 \\
2013	UF15	&	10	&	3:5	&	42.183	&	0.163	&	12.4 &	39.4	&	8.5 	&	5.3   	&	37.3 	& 0.003 \\
2013	TH172	&	4	&	3:5	&	42.458	&	0.261	&	11.6 &	42.6	&	287.7	&	32.6	&	82.3	& 0.039 \\
2013	SU100	&	4	&	3:5	&	42.418	&	0.232	&	9.7	 &	21.5 	&	230.3	&	104.0	&	65.5	& 0.020 \\
2013	RB98	&	3	&	3:5	&	42.120	&	0.255	&	21.9 &	146.9	&	350.8	&	292.6	&	127.7 	& 0.031 \\
2012	UC178	&	7	&	3:5	&	42.143	&	0.168	&	2.5	 &	178.3 	&	168.9	&	59.1	&	132.4	& 0.004 \\
2012	TC324	&	9	&	3:5	&	42.266	&	0.197	&	9.6	 &	131.7	&	258.6	&	6.3	    &	34.7 	& 0.003 \\
2012	BY154	&	11	&	3:5	&	42.190	&	0.168	&	7.1  &	3.7	    &	193.3	&	342.6	&	23.3 	& 0.004 \\
2011	UO411	&	2	&	3:5	&	42.220	&	0.205	&	5.3  &	310.0	&	98.4 	&	360.1 	&	17.3	& 0.005 \\
2011	UN411	&	2	&	3:5	&	42.289	&	0.211	&	3.2	 &	326.2 	&	60.4	&	14.5 	&	48.0 	& 0.010 \\
2011	UM411	&	3	&	3:5	&	42.187	&	0.147	&	9.4	 &	11.3 	&	198.9 	&	192.0 	&	51.2	& 0.008 \\
2011	UL411	&	2	&	3:5	&	42.192	&	0.072	&	2.0  &	308.2 	&	80.3 	&	12.3	&	110.0   & 0.007	\\
2011	UK411	&	9	&	3:5	&	42.242	&	0.231	&	13.5 &	219.3	&	209.9	&	347.0	&	41.0	& 0.003 \\
2011	UJ411	&	2	&	3:5	&	42.231	&	0.117	&	5.1	 &	21.6	&	28.5 	&	357.0	&	30.1 	& 0.005 \\
2008	CS190	&	11	&	3:5	&	42.156	&	0.156	&	16.0 &	326.7	&	150.7	&	37.3	&	52.3	& 0.003 \\
2006	QQ180	&	11	&	3:5	&	42.284	&	0.178	&	9.4	 &	145.5	&	153.4	&	74.4	&	76.7 	& 0.003 \\
2006	CJ69	&	5	&	3:5	&	42.050	&	0.227	&	17.9 &	130.5 	&	60.1	&	342.6	&	75.6	& 0.020 \\
2005	TN74	&	14	&	3:5	&	42.399	&	0.243	&	2.2	 &	179.3	&	224.0	&	357.3 	&	96.1	& 0.006 \\
2005	SE278	&	6	&	3:5	&	42.307	&	0.112	&	6.9  &	95.1	&	268.9	&	27.9	&	38.6	& 0.015 \\
2004	VE131	&	4	&	3:5	&	42.065	&	0.253	&	5.2	 &	219.2	&	272.7 	&	318.5 	&	50.1 	& 0.069 \\
2003	YW179	&	7	&	3:5	&	41.920	&	0.148	&	2.4	 &	97.2	&	18.4	&	19.1	&	116.5 	& 0.013 \\
2003	US292	&	5	&	3:5	&	42.114	&	0.253	&	7.7	 &	29.2	&	346.3	&	21.6	&	81.4	& 0.048 \\
2002	VV130	&	5	&	3:5	&	42.208	&	0.166	&	2.4	 &	171.5 	&	258.8 	&	346.7 	&	28.7 	& 0.042 \\
2002	VA131	&	10	&	3:5	&	42.120	&	0.237	&	7.1	 &	224.5	&	254.9	&	318.8 	&	16.0 	& 0.010 \\
2001	YH140	&	12	&	3:5	&	42.081	&	0.136	&	11.1 &	108.9	&	354.4	&	27.2	&	94.1 	& 0.002 \\
2001	XP254	&	7	&	3:5	&	41.935	&	0.212	&	2.6  &	305.0 	&	181.3 	&	7.1 	&	143.8 	& 0.005 \\
2001	QF331	&	6	&	3:5	&	42.302	&	0.252	&	2.7	 &	156.8	&	248.7	&	349.0	&	76.0 	& 0.009 \\
2000	QN251	&	7	&	3:5	&	42.230	&	0.126	&	0.3	 &	43.0 	&	315.9 	&	22.7 	&	112.7	& 0.029 \\
1999	CX131	&	8	&	3:5	&	42.127	&	0.236	&	9.8	 &	128.1	&	111.1	&	305.4	&	72.37	& 0.027 \\
1994	JS	    &	7	&	3:5	&	42.664	&	0.225	&	14.0 &	56.2 	&	237.4	&	356.1	&	64.8	& 0.036 \\
\\
2015	VZ166	&	4	&	4:7	&	43.579 	&	0.059 	&	4.5	 &	61.1 	&	350.5 	&	3.6	    &	157.6 	& 0.011 \\
2015	VY166	&	3	&	4:7	&	43.564 	&	0.076	&	1.2	 &	60.8 	&	277.8	&	70.5	&	146.4 	& 0.021 \\
2015	VW166	&	3	&	4:7	&	43.525 	&	0.089 	&	1.5	 &	59.2 	&	74.2	&	292.9 	&	125.6	& 0.020 \\
2015	VV166	&	3	&	4:7	&	43.565 	&	0.078	&	2.8	 &	148.3 	&	49.2	&	224.9 	&	136.5 	& 0.014 \\
2015	VU166	&	3	&	4:7	&	43.601	&	0.090	&	3.6	 &	76.5 	&	153.4 	&	188.3	&	122.6	& 0.010 \\
2015	VT166	&	3	&	4:7	&	43.532 	&	0.199 	&	8.3	 &	67.0 	&	72.0 	&	300.9	&	27.5    & 0.042 \\
2015	VS166	&	4	&	4:7	&	43.599	&	0.137	&	2.7	 &	69.7 	&	197.5 	&	136.7	&	125.1 	& 0.015 \\
2015	VR166	&	3	&	4:7	&	43.462	&	0.166	&	6.9	 &	227.2	&	186.3	&	2.6	    &	171.0	& 0.009 \\
2015	VQ166	&	3	&	4:7	&	43.580	&	0.051	&	2.0  &	58.3 	&	272.4 	&	77.7	&	137.4	& 0.010 \\
2015	VC167	&	3	&	4:7	&	43.566	&	0.137 	&	3.7	 &	163.6 	&	278.8 	&	340.3	&	97.6 	& 0.017 \\
2015	TL361	&	4	&	4:7	&	43.705	&	0.165	&	10.6 &	72.8	&	318.3	&	357.0	&	156.7 	& 0.031 \\
2015	RG278	&	3	&	4:7	&	43.769	&	0.172 	&	3.6  &	332.4 	&	26.6 	&	12.7 	&	106.0	& 0.014 \\
2015	RD278	&	3	&	4:7	&	43.789 	&	0.116 	&	2.6  &	192.2	&	173.0 	&	7.8 	&	95.0 	& 0.010 \\
2015	KB174	&	4	&	4:7	&	43.868 	&	0.063 	&	6.0  &	153.6	&	254.1	&	195.4 	&	164.0 	& 0.009 \\
2015	GJ55	&	3	&	4:7	&	43.902 	&	0.171 	&	3.9  &	48.7	&	131.1	&	24.0	&	135.5 	& 0.016 \\
2015	GG55	&	3	&	4:7	&	43.611	&	0.163	&	3.0  &	129.7 	&	52.2	&	20.1 	&	138.1	& 0.017 \\
2015	GF55	&	3	&	4:7	&	43.712	&	0.247 	&	5.3  &	202.0 	&	324.5	&	30.3 	&	50.8	& 0.023 \\
2015	GE55	&	3	&	4:7	&	43.740 	&	0.060 	&	1.8  &	57.9	&	154.0	&	0.0 	&	157.2	& 0.009 \\
2015	GC55	&	3	&	4:7	&	43.764 	&	0.081 	&	2.0  &	199.0 	&	333.7 	&	32.9	&	113.2	& 0.013 \\
2015	FP345	&	8	&	4:7	&	43.768	&	0.217 	&	10.0 &	130.2 	&	34.8	&	31.0	&	45.1 	& 0.004 \\
2015	BP518	&	9	&	4:7	&	43.599 	&	0.177 	&	10.2 &	105.0 	&	0.5	    &	71.0 	&	142.2 	& 0.004 \\
2014	UM229	&	4	&	4:7	&	43.661 	&	0.132 	&	4.2  &	244.8 	&	165.4 	&	348.5 	&	51.4	& 0.010 \\
2014	TL95	&	3	&	4:7	&	43.557	&	0.188	&	10.6 &	100.8	&	306.3	&	352.3	&	124.3 	& 0.026 \\
2014	AL55	&	9	&	4:7	&	43.441	&	0.245	&	4.3  &	104.7 	&	75.6 	&	332.2 	&	51.9	& 0.004 \\
2013	TK187	&	3	&	4:7	&	43.736	&	0.060	&	6.6	 &	56.8	&	296.8	&	27.2	&	151.9	& 0.030 \\
2013	TF187	&	2	&	4:7	&	43.784	&	0.180	&	2.4	 &	80.7	&	250.2	&	33.7	&	55.8	& 0.037 \\
2013	TC187	&	4	&	4:7	&	43.783	&	0.066	&	1.0	 &	72	    &	291.4	&	11.7	&	83.4 	& 0.094\\
2013	SB101	&	5	&	4:7	&	43.730 	&	0.117	&	2.1  &	111.5	&	286.3	&	346.7 	&	118.4	& 0.010 \\
2013	RE109	&	5	&	4:7	&	43.793	&	0.155	&	5.4  &	112.7	&	166.5	&	79.9	&	145.0 	& 0.027 \\
2013	GR136	&	4	&	4:7	&	43.833	&	0.081 	&	1.6  &	198.7 	&	56.1 	&	332.8 	&	90.0	& 0.010 \\
2013	FR28	&	8	&	4:7	&	43.753	&	0.241 	&	3.0  &	41.9	&	133.1 	&	21.22 	&	27.8    & 0.007 \\
2013	BN82	&	11	&	4:7	&	43.342	&	0.204	&	6.6	 &	193.8	&	277.9	&	15.9	&	108.0 	& 0.004 \\
2012	YO9	    &	4	&	4:7	&	43.540	&	0.165	&	15.4 &	118.4	&	3.3	    &	304.6 	&	67.0	& 0.080 \\
2005	SF278	&	13	&	4:7	&	43.604	&	0.185	&	13.4 &	101.5	&	320.4	&	356.6	&	132.9	& 0.003  \\
2004	VF131	&	6	&	4:7	&	43.574	&	0.217	&	0.8  &	246.4 	&	63.3	&	87.1 	&	99.3	& 0.037 \\
2004	OK14	&	5	&	4:7	&	44.198	&	0.249	&	3.5  &	149	    &	206.3	&	343.7	&	127.1 	& 0.016 \\
2003	QW111	&	8	&	4:7	&	43.891 	&	0.110 	&	2.7  &	68.4 	&	17.9	&	283.6	&	109.1	& 0.019  \\
2002	PA149	&	10	&	4:7	&	43.777	&	0.176	&	4.0	 &	105.4	&	154.6	&	88.6	&	151.8	& 0.005 \\
2001	QE298	&	13	&	4:7	&	43.830	&	0.158	&	3.7	 &	7.7	    &	9.7	    &	0.9  	&	96.1 	& 0.003 \\
2001	KP77	&	4	&	4:7	&	44.082 	&	0.181 	&	3.3  &	22.1	&	219.5	&	21.9	&	110.7 	& 0.043 \\
2001	KO76	&	7	&	4:7	&	43.932	&	0.113	&	2.2  &	44.2 	&	295.0	&	276.4 	&	115.1 	& 0.007 \\
2001	KJ76	&	4	&	4:7	&	44.038	&	0.085 	&	6.7  &	47.6 	&	268.5	&	318.0 	&	151.6	& 0.047 \\
2000	OY51	&	6	&	4:7	&	44.014	&	0.237	&	11.2 &	284.9	&	81.0	&	333.6	&	105.2	& 0.012 \\
2000	OP67	&	7	&	4:7	&	43.960 	&	0.193	&	0.8  &	182.4 	&	79.4 	&	67.8	&	64.5 	& 0.029 \\
2000	FD8	    &	6	&	4:7	&	44.022	&	0.226	&	19.5 &	184.8 	&	81.0	&	328.5 	&	141.5 	& 0.003 \\
1999	KR18	&	3	&	4:7	&	43.850	&	0.202	&	0.6  &	283.4	&	213.6 	&	193.0	&	134.0	& 0.073 \\
1999	HT11	&	5	&	4:7	&	44.016	&	0.119 	&	5.0  &	88.0	&	188.1 	&	328.1 	&	152.9 	& 0.006 \\
1999	HR11	&	8	&	4:7	&	43.806	&	0.037 	&	3.3  &	83.4	&	133.7	&	1.4 	&	156.6	& 0.007 \\
1999	HG12	&	6	&	4:7	&	43.881 	&	0.160 	&	1.0 &	30.7 	&	267.0 	&	310.8 	&	67.2 	& 0.155\\
1999	CO153	&	7	&	4:7	&	43.456	&	0.082	&	0.8	&	278.6	&	156.3	&	67.5	&	145.9	& 0.019 \\
\\
2012	HE85	&	5	&	5:9	&	44.936 	&	0.106 	&	3.0 	&	235.0 	&	38.7	&	12.8	&	122.2	& 0.009 \\
2004	VS75	&	6	&	5:9	&	44.483 	&	0.106 	&	7.4	    &	23.6	&	292.7	&	71.2	&	168.4	& 0.047 \\
2002	GD32	&	5	&	5:9	&	44.630 	&	0.133 	&	6.6 	&	22.2	&	23.4 	&	181.0	&	106.0 	& 0.032 \\
2001	KL76	&	4	&	5:9	&	44.811	&	0.094	&	1.3 	&	255.3 	&	139.6	&	235.8 	&	58.3    & 0.056 \\
2000	FR53	&	5	&	5:9	&	44.676 	&	0.083 	&	2.5 	&	39.4 	&	90.9 	&	91.8 	&	164.8	& 0.046 \\
\\
2014	MC70	&	8	&	6:11	&	45.483	&	0.133 	&	7.7	 &	25.6 	&	281.9	&	11.6 	&	152.8	& 0.004 \\
2013	UM15	&	6	&	6:11	&	45.129 	&	0.079	&	1.8  &	110.9	&	183.2 	&	81.2 	&	144.8 	& 0.009 \\
2002	VS130	&	6	&	6:11	&	44.867 	&	0.122	&	3.0  &	275.5	&	86.7 	&	64.0 	&	162.8	& 0.036 \\
2001	KU76	&	6	&	6:11	&	45.348	&	0.169	&	10.6 &	45.0 	&	206.0 	&	358.5 	&	165.7	& 0.031 \\
2010	LQ68$^\star$	&	2	&	6:11	&	45.440 	&	0.231	&	3.2	 &	306.3 	&	10.6 	&	326.7 	&	83.3	& 7.859 \\
1999	CP153$^\star$	&	2	&	6:11	&	44.840 	&	0.162	&	3.0  &	122.9 	&	244.5	&	128.4	&	132.5 	& 8.999 \\
\\
2015	RP278	&   3	&   7:12    &   43.247 	&  0.168  &	 7.2  &	354.8  &	 29.8  &	354.4  & 	122.8   & 0.014 \\
2013	TB172	&   3	&   7:12    &   43.150	&  0.198  &	11.6  &	35.2   &	326.9  &	 13.6  &	163.3   & 0.033 \\
\\
1998	WX31	&   9	&   7:13    &   45.163	&  0.102  &	 3.0  & 37.4   &	40.9   &	17.1   &	118.6   & 0.012 \\
\end{longtable}
\clearpage
\twocolumn

\clearpage
\section{Migration speeds of Neptune's exterior resonances}
\label{sec:migration speed}

Here we supply the migration speeds of two adjacent resonances when they pass through the same radial location one after the other during the outward migration of Neptune.

In the model of a smooth migration, the change of Neptune's semimajor axis $a_N$ can be described by a simple time variation \citep{Malh1995} 
\begin{equation}
  a_N(t)=a_N^{(f)}-\Delta a_N \exp(-t/\tau),
\label{eq:variation}
\end{equation}
where $a_N^{(f)}$ is Neptune's current semimajor axis, $a_N(t)$ is the value at time $t$, $\Delta a_N$ is the amplitude of the migration, and $\tau$ is the migration timescale. 

Since the nominal location of the $p:q$ resonance with Neptune is defined as
\begin{equation}
  a_{res}=a_N\cdot\sqrt[3]{q^2/p^2},
\label{eq:ResLoc}
\end{equation}
where $p<q$ as we used for the high-order resonances in the MCKB. Let $\alpha=\sqrt[3]{q^2/p^2}$, and combining equations (\ref{eq:variation}) and (\ref{eq:ResLoc}), we can obtain the expression for the migration of Neptune's exterior resonance
\begin{equation}
  a_{res}(t)=\alpha a_N^{(f)}-\alpha \Delta a_N \exp(-t/\tau).
\label{eq:ResVar}
\end{equation}
Accordingly, the migration rate of the resonance can be written using the time derivative of $a_{res}$, as
\begin{equation}
  \dot{a}_{res}=\frac{\mbox{d}{a}_{res}}{\mbox{d}{t}}=\frac{\Delta a_N}{\tau}\cdot\alpha\cdot\exp(-t/\tau).
\label{eq:ResRat}
\end{equation}

If we assume that, at the time of $t_{\ast}$, the resonance reaches a specific location of $a_{\ast}$ in the planetesimal disk beyond the orbit of Neptune, i.e.  
\begin{equation}
  a_{res}(t_{\ast})=a_{\ast}~~~~~~~~(a_{\ast}>a_N(t)),
\label{eq:location}
\end{equation}
then we have the following equation
\begin{equation}
  \alpha \Delta a_N \exp(-t_{\ast}/\tau) = \alpha a_N^{(f)} - a_{\ast}.
\label{eq:LocTim}
\end{equation}
Substituting the left-side of equation (\ref{eq:LocTim}) into equation (\ref{eq:ResRat}), we can get the resonance's migration speed at the time of $t_{\ast}$, as
\begin{equation}
  \dot{a}_{res}(t_{\ast})=\alpha a_N^{(f)} - a_{\ast}.
\label{eq:ResRat2}
\end{equation}

Next, we consider the migration speed ratio between two successive resonances when they pass the same location $a_{\ast}$. This ratio can be written as
\begin{equation}
  \frac{\dot{a}_{res}^{A}}{\dot{a}_{res}^{B}}=\frac{\alpha^{A} a_N^{(f)} - a_{\ast}}{\alpha^{B} a_N^{(f)} - a_{\ast}},
\label{eq:RatRatio}
\end{equation}
where the superscript $A$ refers to resonance $A$ that reaches $a_{\ast}$ first, and the superscript $B$ denotes the following resonance $B$. Because resonance $A$ is farther away from Neptune than resonance $B$ (see Fig. \ref{width}), then $\alpha^{A}>\alpha^{B}$. Finally, from equation (\ref{eq:ResRat2}) we obtain the relation
\begin{equation}
  \dot{a}_{res}^{A}>\dot{a}_{res}^{B}.
\label{eq:RatRel}
\end{equation}
So for any pair of resonances, caused by an outward-migrating Neptune, the resonance closer to Neptune will sweep radially across any given point ($a_{\ast}$) at a slower speed than the resonance that is farther from Neptune.  

As an example, considering the 4:7 resonance (i.e. resonance $A$) and 3:5 resonance (i.e. resonance $B$), equation (\ref{eq:RatRel}) proves that the 3:5 sweeps past the same location in the planetesimal disk at a lower speed than the 4:7, as we argued in Sect. 4.1.


\label{lastpage}

\end{document}